\documentclass[journal]{IEEEtran}

\usepackage{cite}
\usepackage{multicol,multirow}
\usepackage{makecell,booktabs}
\usepackage{xurl}
\usepackage{hyperref}
\hypersetup{    
    colorlinks=true,
    linkcolor=black,
    anchorcolor=black,
    citecolor=black,
    filecolor=black,
    menucolor=black,
    runcolor=black,
    urlcolor=black,
}

\usepackage{caption}
\usepackage{subcaption}
\usepackage[utf8]{inputenc} 
\usepackage{graphicx}
\usepackage{amsfonts}
\usepackage{amssymb}
\usepackage{amsmath,mathtools}
\usepackage{array}

\usepackage{color}
\usepackage{svg}
\usepackage{bbm}
\usepackage{bm}
\usepackage{comment}
\usepackage{forest}
\usepackage{etoolbox}
\usepackage{xspace}

\usepackage{cleveref}
\usepackage{lipsum}  
\usepackage{balance}
\usepackage{tabularx}
\usepackage{pifont}

\usepackage{tikz}
\usetikzlibrary{
    arrows.meta,
    positioning,
    decorations.pathreplacing,
    shapes.misc
}

\usepackage{verbatim}

\newcolumntype{L}[1]{>{\raggedright\let\newline\\\arraybackslash\hspace{0pt}}m{#1}}
\newcolumntype{C}[1]{>{\centering\let\newline\\\arraybackslash\hspace{0pt}}m{#1}}
\newcolumntype{R}[1]{>{\raggedleft\let\newline\\\arraybackslash\hspace{0pt}}m{#1}}

\DeclareMathOperator*{\concat}{%
    \mathchoice%
        {\Big\Vert}%
        {\big\Vert}%
        {\Vert}%
        {\Vert}%
}

\newcommand{\etal}{\textit{et al.}}

\begin{document}

\author{
    Minsoo~Kim,~\IEEEmembership{Member,~IEEE,}
    Vladimir~Dvorkin,~\IEEEmembership{Member,~IEEE,}
    and Jip~Kim,~\IEEEmembership{Member,~IEEE}
    }

\title{
    {Probabilistic Dynamic Line Rating with\\Line Graph Convolutional LSTM}
}
\maketitle

\begin{abstract}
Dynamic line rating (DLR) is an effective approach to enhancing the utilization of existing transmission line infrastructure by adapting line ratings according to real-time weather conditions. 
Accurate DLR forecasts are essential for grid operators to effectively schedule generation, manage transmission congestion, and lower operating costs. As renewable generation becomes increasingly variable and weather-dependent, accurate DLR forecasts are also crucial for improving renewable utilization and reducing curtailment during congested periods.
Deterministic forecasts, however, often inadequately represent actual line capacities under uncertain weather conditions, leading to operational risks and costly real-time adjustments. To overcome these limitations, we propose a novel network-wide probabilistic DLR forecasting model that leverages both spatial and temporal information, significantly reducing the operational risks and inefficiencies inherent in deterministic methods. 
Case studies on a synthetic Texas 123-bus system demonstrate that the proposed method not only enhances grid reliability by effectively capturing true DLR values, but also substantially reduces operational costs.
\end{abstract}

\vspace{-6mm}
\section*{Selected Nomenclature}\vspace{-3mm}
\addcontentsline{toc}{section}{Nomenclature}

\subsection{Sets}
\begin{IEEEdescription}[\IEEEusemathlabelsep]
\item[$\mathcal{T}, \mathcal{I}, \mathcal{E}$] \quad\; Time slots, bus and transmission lines
\item[$\mathcal{G}_i, \Omega_i$] \quad\; Generators and neighboring buses of bus $i$
\item[$\mathcal{G}_i^C, \mathcal{G}_i^R$] \quad\; Controllable and renewable generators at bus $i$
\end{IEEEdescription}\vspace{-4mm}

\subsection{Optimization Variables}
\begin{IEEEdescription}[\IEEEusemathlabelsep]
\item[$p_{g,t}^G, \theta_{i,t}$] \quad\; Generator $g$ and voltage angle at bus $i$
\item[$r_{g,t}^+, r_{g,t}^-$] \quad\; Up/down redispatch (real‐time)
\item[${p}_{g,t}^\text{cur,da}, {p}_{g,t}^\text{cur,rt}$] \quad\; Renewable curtailment (day-ahead, real-time)
\end{IEEEdescription}\vspace{-4mm}

\subsection{Parameters}
\begin{IEEEdescription}[\IEEEusemathlabelsep]
\item[$c_{g,1}, c_{g,2}$] \quad\; $1^\text{st}$ and $2^\text{nd}$-order generation cost coefficients
\item[$c_g^+, c_g^-$] \quad\; Up/down redispatch cost
\item[$\underline{P}_g^G,\overline{P}_g^G$] \quad\; Min/max power limit of generator $g$
\item[$\underline{R}_g^G,\overline{R}_g^G$] \quad\; Ramp limits for $g$
\item[$\hat{P}_{ij,t}^L, \overline{P}_{ij}^L$] \quad\; Forecasted and true DLR line limit
\item[$\hat{P}_{i,t}^D, P_{i,t}^D$] \quad\; Forecasted and true load at bus $i$
\item[$\underline{\Theta}_{ij},\overline{\Theta}_{ij}$] \quad\; Min/max angle difference for line $(i,j)$
\item[$P_{g,t}^{G*}$] \quad\; Scheduled dispatch (day-ahead)
\item[$P_{g,t}^{G\dagger}$] \quad\; Previously determined dispatch (real-time)
\item[${P}_{g,t}^{G_r}$] \quad\; Renewable generation of $g$
\end{IEEEdescription}\vspace{-4mm}

\section{Introduction}\label{Sec:Intro}

Traditionally, transmission lines have been operated based on static line rating (SLR), which defines a constant maximum allowable current for each line. 
SLRs are determined using conservative weather assumptions, such as high ambient temperatures and low wind speeds, to ensure safe and reliable operation of the power system. 
However, these conservative assumptions frequently lead to significant underutilization of available transmission line capacity \cite{fernandez2016review}.

Since power systems based on SLR underutilize available transmission capacity, they face challenges in delivering power from generation-rich regions to high demand metropolitan areas. Even when such regions are generating enough power, the fixed ratings under the SLR restrict the amount that can be transferred. In addition, since SLR does not account for real-time conditions, systems based on SLR cannot dynamically adjust transmission capacity. Thus, operations relying on SLR are inefficient when addressing increasing demand and variability of renewable generation. Although constructing new lines could alleviate these problems by providing additional capacity, it takes a long period of time to complete. Thus, dynamic line rating (DLR) has emerged as an alternative to fully utilize existing transmission lines \cite{douglass2019review}. Unlike SLR, DLR adjusts line ratings in real time based on weather conditions, which leads to a cost-effective method to expand grid capacity without additional line construction. 

{
\captionsetup[figure]{font={stretch=0.95}}
\begin{figure}[t]
	\centering
\includegraphics[width=1\columnwidth]{Figures/fig_prob_5.pdf}\vspace{-3mm}
	\caption{\small Example of probabilistic and deterministic DLR forecasting. Deterministic forecasting inevitably contains errors, while probabilistic forecasts provide a range of possible line ratings.\vspace{-5mm}}
	\label{fig:ex_prob_deter}
\end{figure}
}

\begin{table*}[t]
    \centering
    \captionsetup{justification=centering, labelsep=period, font=footnotesize, textfont=sc}
    \caption{Comparison of methods for DLR forecasting. $\triangle$ represents partial consideration.}\vspace{-2mm}
    \label{table:related_works_dlr}
    \begin{tabular}{l|lccc}
        \toprule
        \makecell{Method}
        & \makecell{Model and approach} 
        & \makecell{Probabilistic\\forecasting}
        & \makecell{System\\operation}
        & \makecell{Network-wide\\consideration}\\
        \midrule\midrule
        Bhattarai \etal \cite{bhattarai2018improvement}
          & Numerical weather prediction
          & $\times$
          & $\times$
          & $\times$ \\
        Madadi \etal \cite{madadi2019dynamic}
          & Ornstein--Uhlenbeck process
          & $\times$
          & $\times$
          & $\times$ \\
        Saatloo \etal \cite{saatloo2021hierarchical} 
          & Extreme learning machine
          & $\times$
          & $\times$
          & $\times$ \\
          Ahmadi \etal \cite{ahmadi2023decomposition}
          & Decision tree ensemble model
          & $\times$
          & $\times$
          & $\times$ \\
        Martinez \etal \cite{martinez2024dynamic}
          & Deep learning (LSTM)
          & $\times$
          & $\times$
          & $\times$ \\
        Gao \etal \cite{gao2023day} 
          & Deep learning (LSTM)
          & $\times$
          & \checkmark
          & $\times$ \\
          Phillips \etal \cite{phillips2021forecasting}
        & CFD-based DLR with spatial wind resolution 
        & $\times$ 
        & $\times$ 
        & $\triangle$ \\
          \midrule
        Viafora \etal \cite{viafora2020chance}
          & Gaussian mixture model
          & \checkmark
          & \checkmark
          & $\times$ \\
        Kirilenko \etal \cite{kirilenko2020risk}
          & Super quantile regression
          & \checkmark
          & $\times$
          & $\times$ \\
        Aznarte \etal \cite{aznarte2016dynamic}, Molinar \etal \cite{molinar2019ampacity}
          & Quantile regression forest
          & \checkmark
          & $\times$
          & $\times$ \\
        Dupin \etal \cite{dupin2019optimal}
          & Quantile regression forest
          & \checkmark
          & \checkmark
          & $\times$ \\
        Sun \etal \cite{sun2022spatio}
          & Spatio-temporal regression
          & \checkmark
          & $\times$
          & $\triangle$ \\
        \midrule
        \textbf{LGCLSTM (proposed)}
          & \textbf{Deep learning (line graph convolutional LSTM)}
          & \checkmark
          & \textbf{\checkmark}
          & \checkmark \\
        \bottomrule
    \end{tabular}\vspace{-4mm}
\end{table*}

Meanwhile, the regulatory landscape has also changed to reflect the growing recognition of DLR. For example, FERC Order 881 directs transmission operators to adopt ambient adjusted rating (AAR), which primarily account for real-time ambient temperature \cite{ferc2024dlr}. Although AAR is less accurate than DLR because it does not consider wind speed or solar irradiance, Order 881 lays the groundwork for a wider adoption of DLR. In addition, Order 1920 requires operators to include DLR in long-term transmission planning \cite{ferc2024transmission}. These regulatory changes have encouraged system operators, such as ERCOT, to integrate dynamic ratings into grid operations \cite{ERCOTNodalProtocols}.

Furthermore, due to the real-time fluctuations of DLR, it is crucial to incorporate DLR forecasts into operational scheduling and dispatch. Nonetheless, incorporating accurate DLR into grid operations remains challenging, as deterministic forecasting methods often fail to capture true line capacity, which lead to increased operational risks and inefficiencies. This limitation becomes more significant with the growing penetration of renewable generation, whose variability may increase congestion and can force renewable curtailment when incorrect line ratings are used. Consequently, despite the growing interest in DLR \cite{gao2023day}, these challenges have limited its effectiveness, and developing accurate probabilistic DLR forecasting models is essential to realize its full benefits.

Recently, the development of data-driven algorithms has led to significant breakthroughs in power systems \cite{van2021machine}. These advances have prompted the application of machine learning and deep learning for DLR forecasting. As summarized in Table~\ref{table:related_works_dlr}, early works on deterministic DLR forecasting relied on numerical weather prediction (NWP) techniques \cite{bhattarai2018improvement} with physical heat balance equations to generate deterministic ratings. Subsequent studies \cite{madadi2019dynamic,saatloo2021hierarchical,ahmadi2023decomposition} adopted more advanced machine learning (ML) methods: \cite{madadi2019dynamic} used an Ornstein-Uhlenbeck process on historical DLR data, while \cite{saatloo2021hierarchical} utilized extreme learning machines. Meanwhile, \cite{ahmadi2023decomposition} used decision tree ensemble methods for deterministic DLR forecasting. More recently, LSTM-based approaches have shown promise in modeling the temporal patterns of DLR \cite{martinez2024dynamic, gao2023day}. \cite{phillips2021forecasting} used spatial variation in wind conditions for DLR forecasting.

Although there is a huge potential in DLR forecastings, there are several challenges. \textbf{First}, deterministic forecasting inevitably contains forecasting errors, as illustrated in Fig.~\ref{fig:ex_prob_deter}, which can lead to the risk due to the incorrect forecastings of transmission line capacity. Thus, probabilistic DLR forecasting is essential to deal with uncertain weather conditions. \textbf{Second}, although many previous studies proposed forecasting methods, they often did not address how predicted line ratings integrate into system-level operations (e.g., scheduling, congestion management). Without explicit operational integration, the benefits of DLR cannot be fully realized. \textbf{Third}, existing approaches focus on individual transmission lines without considering spatial correlations and interactions within the network. However, incorporating these network-wide correlations is crucial to improve overall forecast performance \cite{song2024graph}.

Several efforts have been made in the literature to address these challenges. Regarding the first challenge, \cite{viafora2020chance} employed a Gaussian mixture model for probabilistic DLR forecasting based on historical weather and line-rating data. \cite{kirilenko2020risk} adopted super quantile regression, while \cite{sun2022spatio} developed a spatio-temporal regression framework. One of the most widely used technique is quantile regression forests (QRF) \cite{meinshausen2006quantile}, which has been applied in \cite{aznarte2016dynamic, molinar2019ampacity, dupin2019optimal}. However, only \cite{gao2023day, viafora2020chance, dupin2019optimal} addressed the second challenge by examining how forecasted DLRs influence system-level operations.

Despite these advances, existing works typically focus on employing data from a limited subset of transmission lines and nearby weather stations \cite{phillips2021forecasting, sun2022spatio}. Consequently, none of these works fully overcomes the third challenge. 
The authors of \cite{phillips2021forecasting} employed spatial information of wind conditions, but their approach is limited to individual transmission lines without considering network topology. Furthermore.  \cite{sun2022spatio} considered both temporal and spatial aspects for probabilistic forecasting, but it relied solely on local weather information. As a result, these works missed the broader interactions throughout the entire transmission network.

Recent advances in graph neural networks (GNN) offer promising tools to overcome the third challenge \cite{kipf2016semi}. 
By using message passing to aggregate information from neighboring nodes, GNNs can effectively learn spatial correlation across the network. The value of GNNs has been explored in various power system applications, such as learning the AC power flow \cite{liu2022topology}, renewable energy  forecasting \cite{song2024graph}, and other applications \cite{liao2021review}, but their application in DLR remains largely unexplored.

In this regard, we propose a novel DLR forecasting algorithm to overcome the aforementioned three challenges. To deal with the first challenge, our proposed method forecasts the prediction interval of uncertain DLRs based on quantile forecasting \cite{wang2019probabilistic}. For the second challenge, we solve a day-ahead optimal power flow (OPF) using both forecasted DLR and load, and then adjust the errors in real-time by solving redispatch optimization using true DLR and load to compare the operational cost and reliability. Finally, to overcome the third challenge, the proposed method consists of a line graph convolutional network and LSTM to capture both spatial and temporal correlations across the transmission network.

This paper builds on our preliminary work in \cite{kim2024probabilistic} and makes the following key contributions: 

\begin{enumerate}
    \item We propose a novel network-wide probabilistic DLR forecasting framework called line graph convolutional LSTM (LGCLSTM), which embeds a line graph convolutional network into an LSTM. By jointly exploiting extended spatio-temporal dependencies among transmission lines, LGCLSTM captures dynamic variations and broader spatial correlations in a single layer, enabling more accurate and reliable DLR forecasting across large transmission networks. To the best of our knowledge, this is the first approach to fully integrate network-wide spatial structures and temporal dynamics for probabilistic DLR forecasting using GNN.
    \item We analyze the operational impact of probabilistic DLR forecasting by integrating the proposed LGCLSTM into a two-stage grid-operation model consisting of day-ahead scheduling and real-time redispatch. Notably, we demonstrate the necessity of DLR forecasting: although SLR exhibits a similar overestimation frequency to conservative probabilistic limits, it performs substantially worse in grid operation, resulting in higher operational cost and greater renewable curtailment because it fails to capture the temporal variations of true DLR. Moreover, our study shows that grid operation can benefit from selecting an appropriate safety margin for DLR, which is enabled by probabilistic DLR forecasting.
    \item We evaluate LGCLSTM against four probabilistic forecasting algorithms \cite{meinshausen2006quantile, wang2019probabilistic, zhao2019t, simeunovic2021spatio} on the Texas 123-bus backbone system \cite{lu2025synthetic}. In particular, we compare LGCLSTM with two widely adopted DLR forecasting methods \cite{meinshausen2006quantile, gao2023day} to confirm its operational superiority, and we assess the performance across multiple prediction intervals (80\%, 90\%, and 98\%) to reflect diverse risk levels. The results show that LGCLSTM consistently achieves the best forecasting performance and reduces operational cost, while yielding a lower renewable curtailment than SLR, demonstrating its notable ability in DLR forecasting.
\end{enumerate}

\section{Probabilistic DLR Forecasting Model}\label{sec:methodology}

\begin{figure}[t]
	\centering
\includegraphics[width=1\columnwidth]{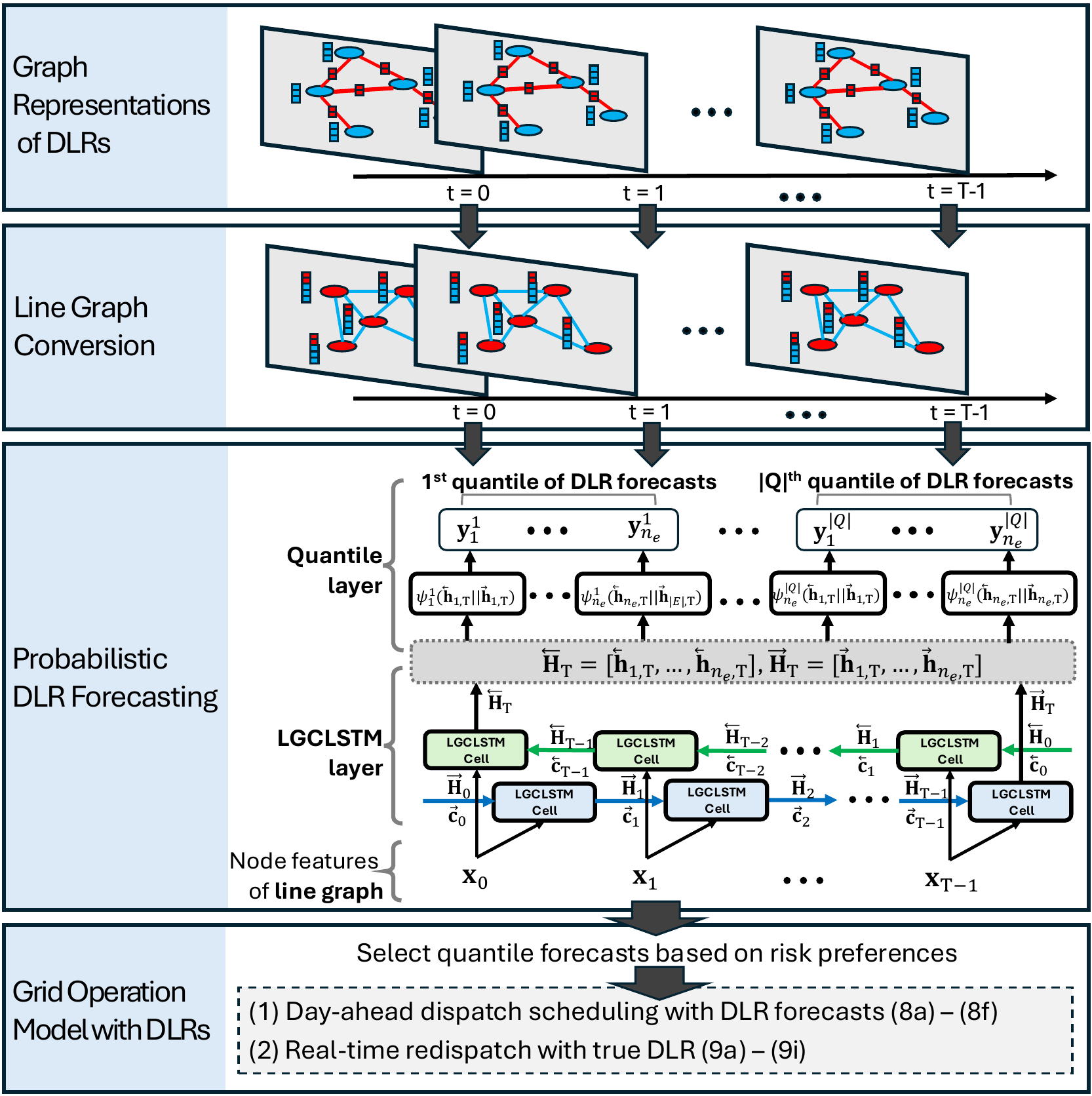}
\vspace{-6mm}
	\caption{\small Overall  framework for probabilistic DLR forecasting.\vspace{-5mm}}
	\label{fig:dlr_framework}
\end{figure}

The overall framework of the proposed probabilistic DLR forecasting is summarized in Fig.~\ref{fig:dlr_framework}. First, the transmission network is converted into a line graph representation to obtain consistent edge-level features. The proposed LGCLSTM then processes this line graph as input to capture extended spatio-temporal dependencies and generate probabilistic DLR forecasts. Finally, these forecasts are incorporated into the grid operation model, which consists of the day-ahead dispatch scheduling and the real-time redispatch, based on the operator’s DLR risk preferences.

\subsection{Limitations of GCN in DLR forecasting.}

Graph convolutional networks (GCNs) learn node features by iteratively aggregating and transforming the features of neighboring nodes. This process is referred to as message passing \cite{gilmer2017neural}. In a power system and DLR setting, message passing indicates that each bus gathers relevant data, such as historical local weather conditions, from its connected neighbors, then merges this neighbor information with its own to refine the bus-level features.

To describe this mathematically, let $G = (V, E)$ denote a graph where $V = \{v_1, ..., v_n\}$ is the set of nodes (buses) and $E\subseteq\{\{a, b\}|a,b\in V, a \neq b\}$ is the set of edges (transmission lines). Let $\mathbf{A}$ be the adjacency matrix of $G$, where $\mathbf{A}_{ij} = 1$ indicates there is a transmission line between the bus $v_i$ and $v_j$. GCNs usually add self-loops by defining $\tilde{\mathbf{A}} = \mathbf{A} + \mathbf{I}$. Then, the diagonal matrix $\tilde{\mathbf{D}}$ is  defined as $\tilde{\mathbf{D}}_{ii} = \sum_{j} \tilde{\mathbf{A}}_{ij}$. Now, we define a degree normalized adjacency as follows:
\begin{equation}
    \hat{\mathbf{A}} = \tilde{\mathbf{D}}^{-\frac{1}{2}}\tilde{\mathbf{A}}\tilde{\mathbf{D}}^{-\frac{1}{2}},
    \label{eq:normalized_adjacency}
\end{equation}
which prevents highly connected buses from dominating the aggregation and stabilizes training by balancing the influence of each neighbor. Then, a single GCN layer transforms a node feature $\mathbf{H}^{(l)}$ at layer $l$ into $\mathbf{H}^{(l+1)}$ as follows:
\begin{equation}
    \mathbf{H}^{(l+1)} = \sigma\big(\hat{\mathbf{A}}\mathbf{H}^{(l)}\mathbf{W}^{(l)}\big),
\end{equation}
where $\mathbf{W}^{(l)}$ is a learnable weight matrix, and $\sigma(\cdot)$ is a nonlinear activation function.

The primary challenges arise from the need to integrate the node and edge features to apply GCN. However, GCN is inherently designed to operate on node features and does not directly utilize edge features \cite{kipf2016semi}. Although GCN variants can incorporate edge attributes \cite{velivckovic2017graph}, they often become impractical for large-scale transmission networks with extensive historical datasets due to the high usage of computational resources \cite{li2021training}. This motivates the use of a line graph representation.

Let $f_V:V\rightarrow\mathbb{R}^{n_v}$ and $f_E:E\rightarrow\mathbb{R}^{n_e}$ map the nodes $v\in V$ and edges $e\in E$ to their feature vector where $n_v$ and $n_e$ are the feature dimensions, respectively. Let $R(v) = \{e\in E|v\in e\}$ be the set of edges incident to $v$. Then, we have
\begin{equation}
    \mathbf{x}_v = \Bigg(\concat_{e\in R(v)}f_E(e) \Bigg)\concat f_V(v),
\end{equation}
where $\mathbf{x}_v\in\mathbb{R}^{|R(v)|n_e + n_v}$ is the result of feature concatenation at $v$. Here, $\concat$ denotes the vector concatenation; for example, if $f_E(e_1)=[a_1,a_2]$, and $f_E(e_2)=[b_1,b_2]$, then
$\concat_{\{e_1, e_2\}} = [a_1,a_2,b_1,b_2].$ Note that $|R(v)|$ varies significantly in power systems.  This variability leads to inconsistency of $\text{dim}(\mathbf{x}_v) = |R(v)|n_e + n_v$ in all $v\in V$. Furthermore, this is problematic for GCNs, which require a fixed feature dimension across all nodes for matrix multiplications and batch processing.

\subsection{Consistent node feature dimensions of a line graph.}
\label{subsec:line_graph_conv}
Alternatively, we concatenate the features of each node on its connected edges. Let $S(e) = \{v\in V|v\in e\}$ be the set of nodes connected by $e\in E$. Then, we have 

\begin{equation}
    \mathbf{x}_e = \Bigg(\concat_{v\in S(e)}f_V(v) \Bigg)\concat f_E(e),
\end{equation}
where $\mathbf{x}_e\in\mathbb{R}^{|S(e)|n_v + n_e}$. Since each transmission line connects exactly two buses in power networks, $|S(e)| = 2$ for all $e\in E$. Thus, $\text{dim}(\mathbf{x}_e) = 2n_v+n_e$ is consistent for all edges. However, we cannot apply standard GCN to learn the concatenated edge features since it can only deal with nodes.

To leverage the consistency of edge feature dimensions, we use the line graph convolutional network (LGCN) as follows: First, we convert the graph $G$ to its line graph $L(G) = (V_L, E_L)$ where each node $u\in V_L$ corresponds to an edge $e\in E$ of the original graph $G$ and $E_L = \{\{u_{e_i}, u_{e_j}\}|u_{e_i}, u_{e_j} \in V_L, e_i, e_j \in E,  e_i\neq  e_j, e_i \cap e_j \neq \emptyset\}$ \cite{harary2018graph}. Thus, using a line graph, we effectively treat each edge in $G$ as a node in $L(G)$.

\subsection{Line Graph Convolution with Extended Neighborhoods}
\label{subsec:line_graph_def}
While a standard GCN (or LGCN) aggregates features only from immediately adjacent nodes, this local-only approach may overlook broader network correlations \cite{zhou2017graph}. To address this limitation, we extend LGCN to capture more distant relationships across the network. 

Let $\mathbf{A}_L$ denote the adjacency matrix of the line graph $L(G)$. We compute $\mathbf{A}_L^k$ to account for the extended neighbor over $k$ hops, where $(\mathbf{A}_L^k)_{ij} \neq 0$ if node $i$ can reach node $j$ via a $k$-hop path \cite[Proposition 1]{zhou2017graph}. Then, a degree normalized adjacency is defined as follows:
\begin{equation}
    \hat{\mathbf{A}}_L^k = ({\mathbf{D}}_L^k)^{-\frac{1}{2}} {\mathbf{A}}_L^k ({\mathbf{D}}_L^k)^{-\frac{1}{2}}, 
\end{equation}
where ${\mathbf{D}}_L^k$ is the diagonal degree matrix of ${\mathbf{A}}_L^k$. A notable benefit of this approach is that a single LGCN layer using $\hat{\mathbf{A}}_L^k$ can replace $k$ stacked single-hop LGCN layers. In doing so, we reduce the overall depth of the model, memory usage, and computational overhead.

\subsection{Embedding Multi-hop LGCNs into LSTM}\label{subsec:prediction}
Now, we propose LGCLSTM by embedding LGCN into LSTM. We use $\hat{\mathbf{A}}_L^k$ for multi-hop LGCN. Let $\mathbf{x}_{u_i,t}$ and $\mathbf{h}_{i,t}$ denote the feature and hidden vector of $i$th node of the line graph $L(G)$ at time $t$, respectively. We have the matrix of feature vectors $\mathbf{X}_{t} = [\mathbf{x}_{u_1, t}, ..., \mathbf{x}_{u_{|E|}, t}]$ and hidden vectors $\mathbf{H}_{t} = [\mathbf{h}_{1,t}, ..., \mathbf{h}_{|E|,t}]$. LGCLSTM cell at time $t$ consists of the forget gate $\mathbf{f}_t$, the input gate $\mathbf{i}_t$, the output gate $\mathbf{o}_t$ and the candidate cell state gate $\mathbf{g}_t$. Then, the hidden state $\mathbf{H}_t$ and cell state $\mathbf{c}_t$ are updated as follows:
\begingroup
\allowdisplaybreaks
\begin{subequations}
	\begin{align}
		&\mathbf{f}_t = \sigma(\hat{\mathbf{A}}_L^k\mathbf{X}_{t-1}\mathbf{W}_f + \mathbf{H}_{t-1}\mathbf{U}_f + \mathbf{b}_f),\\
		&\mathbf{i}_t = \sigma(\hat{\mathbf{A}}_L^k\mathbf{X}_{t-1}\mathbf{W}_i + \mathbf{H}_{t-1}\mathbf{U}_i + \mathbf{b}_i),\\
		&\mathbf{o}_t = \sigma(\hat{\mathbf{A}}_L^k\mathbf{X}_{t-1}\mathbf{W}_o + \mathbf{H}_{t-1}\mathbf{U}_o + \mathbf{b}_o),\\
		&\mathbf{g}_t = \tanh(\hat{\mathbf{A}}_L^k\mathbf{X}_{t-1}\mathbf{W}_g + \mathbf{H}_{t-1}\mathbf{U}_g + \mathbf{b}_g),\\
		&\mathbf{c}_t = \mathbf{f}_t \odot \mathbf{c}_{t-1} + \mathbf{i}_t \odot \mathbf{g}_t,\\
		&\mathbf{H}_t = \mathbf{o}_t \odot \tanh(\mathbf{c}_t),
	\end{align}\label{eq:D-LGCN}
\end{subequations}\vspace{0mm}
\endgroup
\noindent
where $\mathbf{W}_f$, $\mathbf{W}_i$, $\mathbf{W}_o$, and $\mathbf{W}_g$ are learnable weight matrices associated with the input features, $\mathbf{U}_f$, $\mathbf{U}_i$, $\mathbf{U}_o$, and $\mathbf{U}_g$ are learnable weight matrices of hidden state, and $\mathbf{b}_f$, $\mathbf{b}_i$, $\mathbf{b}_o$, and $\mathbf{b}_g$ are learnable biases. $\sigma$ is the sigmoid function and $\odot$ is the element-wise product. Note that we only substitute the input sequence part of LSTM and left the hidden state part as it was to avoid oversmoothing due to repeatedly applying graph deep learning to hidden vectors \cite{chen2020measuring}. Also, we use a bidirectional approach to capture spatio-temporal patterns from both the past and the present. Thus, $\overrightarrow{\mathbf{H}}_t$ and $\overleftarrow{\mathbf{H}}_t$ in Fig.~\ref{fig:dlr_framework} denote the hidden matrices of the forward and backward LGCLSTM cell at time $t$, which capture spatial patterns via $\hat{\mathbf{A}}_L^k$.

\subsection{Quantile Layer for Probabilistic Forecasting}
\label{subsec:quantile_forecast}

For probabilistic forecasting, we use $\psi_i^q$, where $q \in \mathcal{Q}$ denotes the set of all quantile used in training. Each $\psi_i^q$ is a neural network layer that maps the output of LGCLSTM in Fig.~\ref{fig:dlr_framework} to the $q$th quantile forecast of line $i$. Specifically, for any $q \in \mathcal{Q}$, we have $\hat{\mathbf{y}}_i^q = \psi_i^q(\overleftarrow{\mathbf{h}}_{i,T}||\overrightarrow{\mathbf{h}}_{i,T})$. Here, $\overleftarrow{\mathbf{h}}_{i,T}$ and $\overrightarrow{\mathbf{h}}_{i,T}$ are the backward and forward hidden states at the final time $T$. Let $Q_q\in[0,1]$ denote the corresponding quantile level of $q$. Then, the quantile loss function for the line $i$ is \cite{wang2019probabilistic}
\begin{equation}
	\mathcal{L}(\hat{y}_{i,t}^q, y_{i,t}) = \begin{cases}
		Q_q(y_{i,t} - \hat{y}_{i,t}^q), \;\;\quad\quad\quad \hat{y}_{i,t}^q\leq y_{i,t},\\
		(1-Q_q)(\hat{y}_{i,t}^q - y_{i,t}), \quad otherwise,
	\end{cases}\label{eq:quantile_loss}
\end{equation}
where $\hat{y}_{i,t}^q$ and $y_{i,t}$ are $t$th element of $\hat{\mathbf{y}}_i^q$ and true DLR $\mathbf{y}_i$, respectively. Finally, we use $\sum_{q=1}^{|\mathcal{Q}|}\sum_{i=1}^{|E|}\sum_{t=0}^{T-1}\mathcal{L}(\hat{y}_{i,t}^q, y_{i,t})$ to train the model for all $q$ and the prediction horizon $T$.

\section{Grid Operation Model with DLRs}\label{sec:problem_formulation}

To evaluate the effectiveness of DLR forecasts from a grid operation perspective, we adopt an energy-only market framework~\cite{pritchard2010single}. Since generators in this framework earn revenue only for their produced energy, any variations in the line rating directly translate into changes in operational costs.

Specifically, we formulate a day-ahead dispatch scheduling and real-time redispatch problem for the grid operation model. The day-ahead problem determines the dispatch schedules using the DLR forecasts. Once the day-ahead schedules are set based on the DLR forecasts, real-time conditions may differ from these forecasts. Thus, we solve the real-time redispatch problem to ensure that the system remains secure and economically efficient under true conditions. Also, we assume that the system operator obtains high-fidelity estimates of DLR and load (e.g., true values) when solving the real-time problem. By analyzing the cost of these two problems, we can assess how well the forecasting approach supports the cost-effective system operations.

Let $\Xi^\text{DA}\coloneqq\{p_{g,t}^G, p_{g,t}^\text{cur,da},\theta_{i,t}\}$ be the set of optimization variables where $p_{g,t}^G$, $p_{g,t}^\text{cur,da}$, and $\theta_{i,t}$ denote generation, renewable curtailment, and voltage phase angle in day-ahead stage. Let $\mathcal{G}_i$ be the generators at bus $i$, where $\mathcal{G}_i^C$ and $\mathcal{G}_i^R$ denote the controllable and renewable generators at bus $i$.  Then, the day-ahead dispatch with DLR forecasts is formulated as follows:
\begingroup
\allowdisplaybreaks
\begin{subequations}\label{Eq:DispatchScheduling}
\begin{align}
&
\min_{\Xi^\text{DA}}
\sum_{t\in\mathcal{T}} 
\sum_{i\in\mathcal{I}}
\sum_{g\in\mathcal{G}_{i}}
  c_{g,2}\bigl(p_{g,t}^G\bigr)^2 
  + c_{g,1}p_{g,t}^G
\label{Eq:Obj}
\\
&
\underline{P}_{g}^G
\le
p_{g,t}^G
\le
\overline{P}_{g}^G,\quad
\forall i\in\mathcal{I},
\forall g\in\mathcal{G}_{i}^C,
\forall t\in\mathcal{T}
\label{Eq:ControllableBounds}
\\
&
0
\le
p_{g,t}^\text{cur,da}
\le
P_{g,t}^{G_r},\quad
\forall i\in\mathcal{I},
\forall g\in\mathcal{G}_{i}^R,
\forall t\in\mathcal{T}
\label{Eq:RenewableBounds}
\\
&
\underline{R}_g^G 
\le
p_{g,t}^G
-
p_{g,t-1}^G
\le
\overline{R}_g^G,\quad
\forall i\in\mathcal{I}, 
\forall g\in\mathcal{G}_{i}^C, 
\forall t\in\mathcal{T}
\label{Eq:Ramping}
\\
&
-\hat{P}_{ij,t}^L 
\le
B_{ij}\bigl(\theta_{i,t}-\theta_{j,t}\bigr)
\le
\hat{P}_{ij,t}^L,
\forall (i,j)\in\mathcal{E},\forall t\in\mathcal{T}
\label{Eq:LineFlow}
\\
&
\Big(\sum_{g\in\mathcal{G}_i^C} p_{g,t}^G
+
\sum_{g\in\mathcal{G}_i^R} P_{g,t}^{G_r} - p_{g,t}^\text{cur,da}\Big)
-
\hat{P}_{i,t}^{D}\notag\\
&\quad\quad\quad\quad\quad\quad=
\sum_{j\in\Omega_i}
B_{ij}\bigl(\theta_{i,t}-\theta_{j,t}\bigr),
\forall i\in\mathcal{I},\forall t\in\mathcal{T}
\label{Eq:PowerBalance}
\end{align}
\end{subequations}
\endgroup
\noindent
Here, $c_{g,1}$ and $c_{g,2}$ in \eqref{Eq:Obj} are the $1^\text{st}$ and $2^\text{nd}$-order generation coefficients of the generator $g\in\mathcal{G}_i$. 
In \eqref{Eq:ControllableBounds}, the controllable generators are bounded by
$\underline{P}_g^G$ and $\overline{P}_g^G$. 
\eqref{Eq:RenewableBounds} imposes curtailment to the renewable generation ${P}_{g,t}^{G_r}$. The ramping constraints \eqref{Eq:Ramping} apply only to controllable units, ensuring that hourly changes in output do not exceed $\overline{R}_g^G$ or drop below $\underline{R}_g^G$. Line flows are restricted by \eqref{Eq:LineFlow}, where $\hat{P}_{ij,t}^L$ are the forecasted DLRs. When employing the probabilistic forecasting model, one of the predicted quantiles is used as  $\hat{P}_{ij,t}^L$ based on the operator’s risk tolerance under DLR uncertainty. Finally, \eqref{Eq:PowerBalance} enforces the power balance of each bus~$i$. Here, $\Omega_i$ is the set of buses directly connected to $i$, and $B_{ij}$ is the line susceptance.

Next, let $\Xi^\text{RT} \coloneqq \{r_{g,t}^+, r_{g,t}^-, {p}_{g,t}^\text{cur,rt}, \theta_{i,t}\}$ be the set of optimization variables where $r_{g,t}^+$ and $r_{g,t}^-$ are the upward and downward redispatch. ${p}_{g,t}^\text{cur,rt}$ is the renewable curtailment in real-time stage. Then, the real-time redispatch problem at $t$ is:\vspace{0mm}
\begingroup
\allowdisplaybreaks
\begin{subequations}\label{eq:Redispatch}
\begin{align}
&\min_{\Xi^\text{RT}}
\sum_{i\in\mathcal{I}} \sum_{g\in\mathcal{G}_i}
  c_{g,2}\bigl(P_{g,t}^{G*}+r_{g,t}^+ - r_{g,t}^-\bigr)^2\notag\\
  &\quad\quad\quad\quad\quad\quad + c_{g,1}P_{g,t}^{G*}
  + c_g^+ r_{g,t}^+
  - c_g^- r_{g,t}^-
\label{eq:rd_obj}
\\
&0 \le r_{g,t}^+ \le \overline{P}_g^G - P_{g,t}^{G*},
\quad \forall i\in\mathcal{I},g\in\mathcal{G}_i^C
\label{eq:rd_upreserve}
\\
&0 \le r_{g,t}^- \le P_{g,t}^{G*} - \underline{P}_g^G,
\quad \forall i\in\mathcal{I},g\in\mathcal{G}_i^C
\label{eq:rd_downreserve}
\\
&r_{g,t}^+ = 0,\; r_{g,t}^- = 0,
\quad \forall i\in\mathcal{I},g\in\mathcal{G}_i^R
\label{eq:rd_renewables}
\\
&0 \le p_{g,t}^\text{cur,rt} \le P_{g,t}^{G_r},
\quad \forall i\in\mathcal{I},g\in\mathcal{G}_i^R
\label{eq:rt_renewable_bound}
\\
&\underline{P}_g^G
\le P_{g,t}^{G*}+r_{g,t}^+ - r_{g,t}^- 
\le \overline{P}_g^G,
\quad \forall i\in\mathcal{I}, g\in\mathcal{G}_i^C
\label{eq:rd_gen_limits}
\\
&\underline{R}_g^G 
\le \bigl(P_{g,t}^{G*}+r_{g,t}^+ - r_{g,t}^- \bigr)
     - P_{g,t-1}^{G\dagger}
\le \overline{R}_g^G, \forall i\in\mathcal{I}, g\in\mathcal{G}_i^C
\label{eq:rd_ramp}
\\
&-\overline{P}_{ij,t}^L
\le B_{ij}(\theta_{i,t}-\theta_{j,t})
\le \overline{P}_{ij,t}^L,
\quad \forall (i,j)\in\mathcal{E}
\label{eq:rd_flow_limits}
\\
&\Big(\sum_{g\in\mathcal{G}_i^C}
   \bigl(P_{g,t}^{G*} + r_{g,t}^+ - r_{g,t}^-\bigr)
 + \sum_{g\in\mathcal{G}_i^R} P_{g,t}^{G_r} - p_{g,t}^\text{cur,rt}\Big)
 - P_{i,t}^{D}\notag\\
&\quad\quad\quad\quad\quad\quad\quad\quad
=
\sum_{j\in\Omega_i}
  B_{ij}(\theta_{i,t}-\theta_{j,t}),
\quad \forall i\in\mathcal{I}
\label{eq:rd_power_balance}
\end{align}
\end{subequations}
\endgroup
\noindent Here, \eqref{eq:rd_obj} reduces the redispatch cost, where $P_{g,t}^{G*}$  is the previously scheduled dispatch and $r_{g,t}^+$, $r_{g,t}^-$ denote upward and downward redispatch. $c_g^+$ and $c_g^-$ are the corresponding adjustment prices. Note that we have $c_g^+ > c_{g,1} > c_g^-$, since we follow the energy-only redispatch model in \cite{dvorkin2025regression}. \eqref{eq:rd_upreserve}--\eqref{eq:rd_downreserve} limit $r_{g,t}^+$ and $r_{g,t}^-$ for controllable generators, while renewable generators are not redispatched, as specified in \eqref{eq:rd_renewables}.  \eqref{eq:rd_gen_limits} restricts the adjusted dispatch. The ramping constraints in \eqref{eq:rd_ramp} restrict the difference between the adjusted dispatch and the final dispatch at the previous time $P_{g,t-1}^{G\dagger}$. \eqref{eq:rt_renewable_bound} defines the curtailment in the real-time stage. Line flows are restricted by \eqref{eq:rd_flow_limits}.  Finally, \eqref{eq:rd_power_balance} ensures power balance.

\section{Case Studies}\label{sec:case_study}
\subsection{Simulation Settings}
{\subsubsection{Data Preparation}
We use the Texas 123-bus backbone transmission (TX-123BT) system \cite{lu2025synthetic} to verify the performance of the proposed method in probabilistic DLR forecasting. This system contains 123 buses and 255 lines. For DLR forecasting, we reduce the number of lines by merging parallel lines. We use five years of historical weather data for each bus and DLR data for each line from 2017 to 2021, with a one-hour resolution. Weather data include temperature, wind speed, wind direction, and solar radiation, while DLR data consist of line ratings calculated based on the heat balance equation \cite{IEEE738_2012}. Although the DLR values in this study are calculated from the heat balance equation, the proposed framework can be applied directly when real DLR sensing data are available.

We split the dataset into a training and testing set using a 4:1 ratio. Hyperparameters, such as layers, are determined by dividing the training set into training and validation subsets at a 3:1 ratio and selecting the configuration that minimizes the validation loss. After tuning, both training and validation sets are combined, and each model is trained from scratch for 60 epochs. T-GCN is trained with 30 epochs due to large number of parameters. LSTM is trained with 20 epochs since the model learns deterministic forecasting rapidly. We use AdamW optimizer \cite{loshchilov2018decoupled} with learning rate of 0.001. The dimension of the hidden vector is 128 for all the methods except T-GCN, which is 32. The batchsize of all the method are 64. We set $k = 5$ since it shows the smallest quantile score during validation. Each bus includes the previous seven days of historical weather data and its geographical coordinates, whereas each line includes the previous seven days of DLR data, its length, and the current season. The model uses these input data to forecast the network-wide DLR of the next day.

The up and down redispatch prices are $c_g^+ = 3c_{g,1}$ and $c_{g}^- = 0.5c_{g,1}$, which is consistent with \cite{dvorkin2025regression}. Load and generator data are rescaled to ensure the feasibility of the grid operation.
}

{
\captionsetup[figure]{font={stretch=0.95}}
\begin{figure}[t]
	\centering
\includegraphics[width=1\columnwidth]{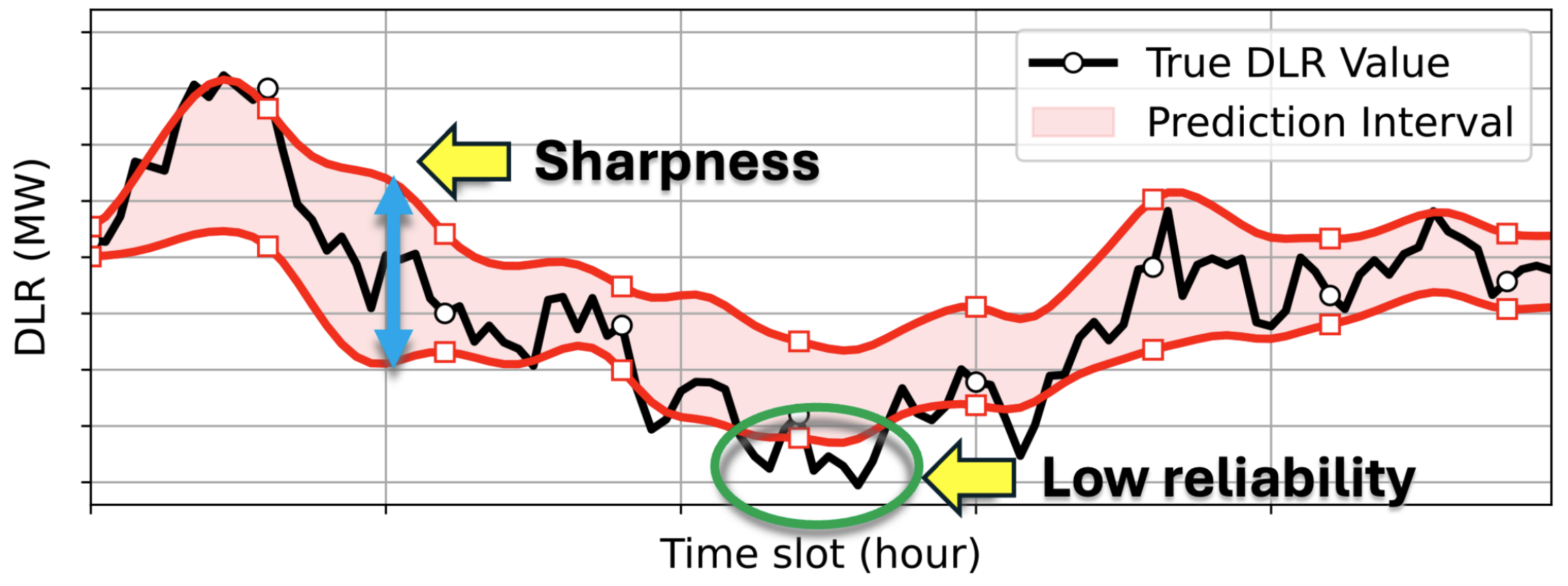}
\vspace{-5.5mm}
	\caption{\small An example of reliability and sharpness. 
    High reliability ensures the prediction intervals consistently capture the true rating, while low sharpness keeps those intervals narrow enough to enable precise capacity utilization in grid operations.\vspace{-2mm}}
	\label{fig:eval_metric}
\end{figure}
}

\subsubsection{Evaluation Metrics}
{
We employ four evaluation metrics to assess the probabilistic forecasting performance, reflecting \textit{reliability} and \textit{sharpness}, as illustrated in Fig.~\ref{fig:eval_metric}. 
In probabilistic DLR forecasting, reliability measures how well the prediction intervals capture the true line ratings, while sharpness reflects the narrowness of those intervals. 
A reliable model ensures that the true DLR values $y_{i,t}$ fall within the predicted interval $[\hat{y}_{i,t}^{L}, \hat{y}_{i,t}^{U}]$, whereas a sharper interval provides higher operational precision but risks missing true values if it becomes too narrow.

\paragraph*{(1) Average Coverage Error (ACE)}
ACE evaluates the deviation between the empirical coverage probability and the nominal confidence level $1-\alpha$ (e.g., $\alpha = 0.1$ for the 90\% prediction interval) \cite{li2022integrated}: 
\begin{equation}
\mathrm{ACE} = 
\Big|
\frac{1}{|E|T} 
\sum_{i=1}^{|E|} \sum_{t=0}^{T-1} \mathbb{I}\!\left( \hat{y}_{i,t}^{L} \le y_{i,t} \le \hat{y}_{i,t}^{U} \right)
- (1-\alpha)
\Big|,
\end{equation}
where $|E|$ is the number of lines, $T$ is the number of time steps, and $\mathbb{I}(\cdot)$ is the indicator function.
A smaller ACE indicates higher reliability.

\paragraph*{(2) Prediction Interval Normalized Average Width (PINAW)}
PINAW measures the average width of the prediction intervals. A lower value indicates sharper intervals \cite{li2022integrated}:
\begin{equation}
\mathrm{PINAW} =
\frac{1}{|E|T}
\sum_{i=1}^{|E|}\sum_{t=0}^{T-1}
\big(\hat{y}_{i,t}^{U} - \hat{y}_{i,t}^{L}\big).
\end{equation}

\paragraph*{(3) Interval Score (IS)}
IS evaluates both reliability and sharpness using the following form \cite{song2024graph}:
\begin{equation}
\begin{aligned}
&\mathrm{IS}_{i,t}^{\alpha} =\\
&\begin{cases}
-2\alpha\,(\hat{y}_{i,t}^{U}-\hat{y}_{i,t}^{L})
-4\,(\hat{y}_{i,t}^{L} - y_{i,t}), & \text{if } y_{i,t} < \hat{y}_{i,t}^{L},\\[3pt]
-2\alpha\,(\hat{y}_{i,t}^{U}-\hat{y}_{i,t}^{L}), & \text{if } \hat{y}_{i,t}^{L} \le y_{i,t} \le \hat{y}_{i,t}^{U},\\[3pt]
-2\alpha\,(\hat{y}_{i,t}^{U}-\hat{y}_{i,t}^{L})
-4\,(y_{i,t} - \hat{y}_{i,t}^{U}), & \text{if } y_{i,t} > \hat{y}_{i,t}^{U}.
\end{cases}
\end{aligned}
\label{eq:is_point}
\end{equation}
\begin{equation}
\mathrm{IS} =
\frac{1}{|E|T}
\sum_{i=1}^{|E|}\sum_{t=0}^{T-1}
\mathrm{IS}_{i,t}^{\alpha}.
\label{eq:is_total}
\end{equation}
A higher $\mathrm{IS}$ indicates a better balance between reliability and sharpness.

\paragraph*{(4) Quantile Score (QS)}
QS evaluates the accuracy of quantile forecasts by minimizing the pinball loss \cite{wang2019probabilistic}:
\begin{equation}
\begin{aligned}
    &\mathrm{QS}=\\
&\frac{1}{|E|T|}
\sum_{i=1}^{|E|}\sum_{t=0}^{T-1}
\begin{cases}
Q_q\,(y_{i,t} - \hat{y}_{i,t}^{q}), & \hat{y}_{i,t}^{q} \le y_{i,t}, \\[3pt]
(1-Q_q)\,(\hat{y}_{i,t}^{q} - y_{i,t}), & otherwise,
\end{cases}
\end{aligned}
\end{equation}
A smaller QS indicates more accurate quantile estimation across the distribution.

Although QS directly measures the accuracy of individual quantile forecasts, it implicitly reflects both reliability and sharpness. 
A reliable forecasting model produces quantile estimates whose empirical coverage matches their nominal levels, 
while a sharper predictive distribution yields smaller quantile deviations from the true values. 

In summary, for ACE, PINAW, and QS, lower values correspond to better performance, whereas a higher IS indicates a better balance between reliability and sharpness.  All metrics are normalized for comparability across test cases and are represented as percentages.
}

\begin{table}[t]
	\centering
    \captionsetup{justification=centering, labelsep=period, font=footnotesize, textfont=sc}
	\caption{Comparison of the baselines methods. \vspace{-2mm}}
	\label{table:baseline_methods}
	\begin{tabular}{c|ccccc}
		\toprule
		\textbf{Method} & \textbf{\makecell{Prob.\\Forecast}}& \textbf{\makecell{Consider\\Network}} & \textbf{\makecell{Line\\Graph}} & \textbf{\makecell{Hops}} & \textbf{\makecell{Layers}} \\
		\midrule\midrule
        LSTM \cite{gao2023day} & $\times$ & $\times$ & $\times$ & -- & 1 \\\midrule
        QRF \cite{meinshausen2006quantile} & \checkmark & $\times$ & $\times$ & -- & -- \\
		QLSTM \cite{wang2019probabilistic} & \checkmark & $\times$ & $\times$ & -- & 1 \\		
        T-GCN \cite{zhao2019t} & \checkmark & \checkmark & $\times$ & Single & 1 \\
		GCLSTM \cite{simeunovic2021spatio} & \checkmark & \checkmark & $\times$ & Single & 3 \\
        \midrule
		$\text{LGCLSTM}$ & \checkmark & \checkmark & \checkmark & Multi & 1 \\
		\bottomrule
	\end{tabular}
\end{table}

\begin{table}[t]
	\centering
    \captionsetup{justification=centering, labelsep=period, font=footnotesize, textfont=sc}
	\caption{The number of learnable parameters of the NN based methods for probabilistic DLR forecastings.}
	\label{table:param_numbers}
	\begin{tabular}{cccc}
		\toprule
        \makecell{\textbf{QLSTM} \\ ($\times 10^7$)} & \makecell{\textbf{T-GCN} \\ ($\times 10^7$)} & \makecell{\textbf{GCLSTM} \\ ($\times 10^7$)} & \makecell{$\textbf{LGCLSTM}$ \\ ($\times 10^7$)} \\
		\midrule\midrule
        6.36 & 25.10 & 3.00 & 1.84\\
		\bottomrule
	\end{tabular}\vspace{-3mm}
\end{table}

\subsubsection{Baseline Models}
We compare the proposed LGCLSTM against five baselines in Table~\ref{table:baseline_methods}. LSTM \cite{gao2023day} focuses solely on the temporal patterns of a single line. QRF \cite{meinshausen2006quantile} is the most widely used method for probabilistic DLR forecasting. Both QRF and QLSTM \cite{wang2019probabilistic} capture only the temporal patterns of a single line. T-GCN \cite{zhao2019t} combines GCN and LSTM sequentially but does not integrate the GCN into the LSTM cell. In contrast, GCLSTM \cite{simeunovic2021spatio} integrates the GCN directly into the LSTM cell. Both T-GCN and GCLSTM operate on the original graph without transforming it into a line graph. LGCLSTM advances further by aggregating features from extended neighborhoods in the line graph. To provide a fair comparison in model efficiency, Table~\ref{table:param_numbers} summarizes the number of learnable parameters for all deep learning-based methods. Notably, LGCLSTM achieves the lowest parameter count while still outperforming all baselines.

\subsection{Forecasting Results}\label{sec:results}

{
\subsubsection{Overall Performance Comparisons}\label{sec:overall}

Table~\ref{table:simulation_results} shows the probabilistic forecasting performance under the 80\%, 90\%, and 98\% prediction intervals. Note that $Q_L$ and $Q_U$ represent the lower and upper bound for each prediction interval, respectively. Across all prediction intervals, LGCLSTM provides the most balanced forecasts, consistently achieving the lowest PINAW and QS and the highest IS. These results indicate that LGCLSTM generates sharp and accurate prediction intervals while maintaining competitive reliability.

At the 80\% interval, QRF attains the smallest ACE, which reflects high reliability, but its substantially larger PINAW reveals excessive conservatism that limits effective DLR utilization. QLSTM exhibits higher ACE than the graph-based baselines. Since QLSTM does not incorporate spatial information, it relies solely on single-line temporal patterns, which leads to an inaccurate representation of the true uncertainty. Thus, QLSTM fails to provide reliable quantile estimates. T-GCN and GCLSTM achieve moderate performance, but the proposed LGCLSTM demonstrates the best balance between reliability and sharpness without excessive conservatism, enabling grid operators to make more informed and cost-effective dispatch decisions.

A similar pattern holds at the 90\% and 98\% intervals. QRF maintains strong reliability but becomes increasingly conservative as the confidence level widens, whereas QLSTM shows a poor balance between sharpness and reliability, as indicated by its smaller IS and larger QS values than graph-based methods. T-GCN and GCLSTM deliver more stable behavior but remain inferior to LGCLSTM in both sharpness and quantile accuracy. LGCLSTM consistently achieves the lowest PINAW and QS and the highest IS across all intervals, confirming its robustness under varying risk levels. This balanced  behavior directly supports the operational benefits discussed in Section~\ref{subsec:opertion_results}.

\begin{table}[t]
	\centering
    \captionsetup{justification=centering, labelsep=period, font=footnotesize, textfont=sc}
	\caption{Performance of probabilistic DLR forecasting methods. The best results are in \textbf{bold}. Lower is better for ACE, PINAW, and QS; higher is better for IS. \vspace{-2mm}}
	\label{table:simulation_results}
	\begin{tabular}{c|l|cccc}
		\toprule
		\multicolumn{2}{c|}{\textbf{Method}}& \textbf{\makecell{ACE\\(\%)}} & \textbf{\makecell{PINAW\\(\%)}} & \textbf{\makecell{IS\\(\%)}} & \textbf{\makecell{QS\\(\%)}}\\
		\midrule\midrule
        \multirow{5}{*}{\makecell{Prediction interval\\of 80\%\\($Q_L = 0.1$\\$Q_U = 0.9$)}}
        &QRF & \textbf{1.37} & 30.80 & -18.23 & 2.28   \\
		&QLSTM                  & 4.48 & 20.19 & -13.50 & 1.69 \\
		&T-GCN                    & 3.45 & 20.74 & -13.25 & 1.66  \\
		&GCLSTM          & 2.10 & 21.76 & -13.61 & 1.70  \\\cmidrule{2-6}
        &$\text{LGCLSTM}$      & 2.16 & \textbf{19.97} & \textbf{-12.70} & \textbf{1.58}   \\
		\midrule\midrule
        \multirow{5}{*}{\makecell{Prediction interval\\of 90\%\\($Q_L = 0.05$\\$Q_U = 0.95$)}}
        &QRF & \textbf{1.63} & 40.55 & -11.25 & 2.42   \\
		&QLSTM   & 4.95 & 26.26 & -8.55 & 1.73   \\
		&T-GCN    & 3.57 & 27.23 & -8.21 & 1.71  \\
		&GCLSTM & 2.44 & 28.66 & -8.45 & 1.77   \\
        \cmidrule{2-6}
        &$\text{LGCLSTM}$      & 2.59 & \textbf{26.13} & \textbf{-7.93} & \textbf{1.64}\\
		\midrule\midrule
        \multirow{5}{*}{\makecell{Prediction interval\\of 98\%\\($Q_L = 0.01$\\$Q_U = 0.99$)}}
        &QRF & 2.41 & 56.32 & -3.35 & 2.95  \\
		&QLSTM                     & 3.59 & 39.07 & -2.76 & 2.10   \\
		&T-GCN 
        & 2.01 & 41.64 & -2.42 & 2.18  \\
		&GCLSTM                  & \textbf{1.72} & 43.44 & -2.48 & 2.26   \\
        \cmidrule{2-6}
        &$\text{LGCLSTM}$      &  2.12 & \textbf{38.78} & \textbf{-2.41} & \textbf{2.05}   \\
        \bottomrule
	\end{tabular}
    {
    \captionsetup{justification=raggedright, singlelinecheck=false, font=footnotesize, textfont=normalfont}
    \caption*{ACE: Reliability score, PINAW: Sharpness score, \\IS: Reliability and sharpness score, QS: quantile accuracy.}
    }\vspace{-2mm}
\end{table}

\begin{figure}[t]
     \centering
     \begin{subfigure}[b]{0.45\columnwidth}
     \centering
         \includegraphics[width=\columnwidth]{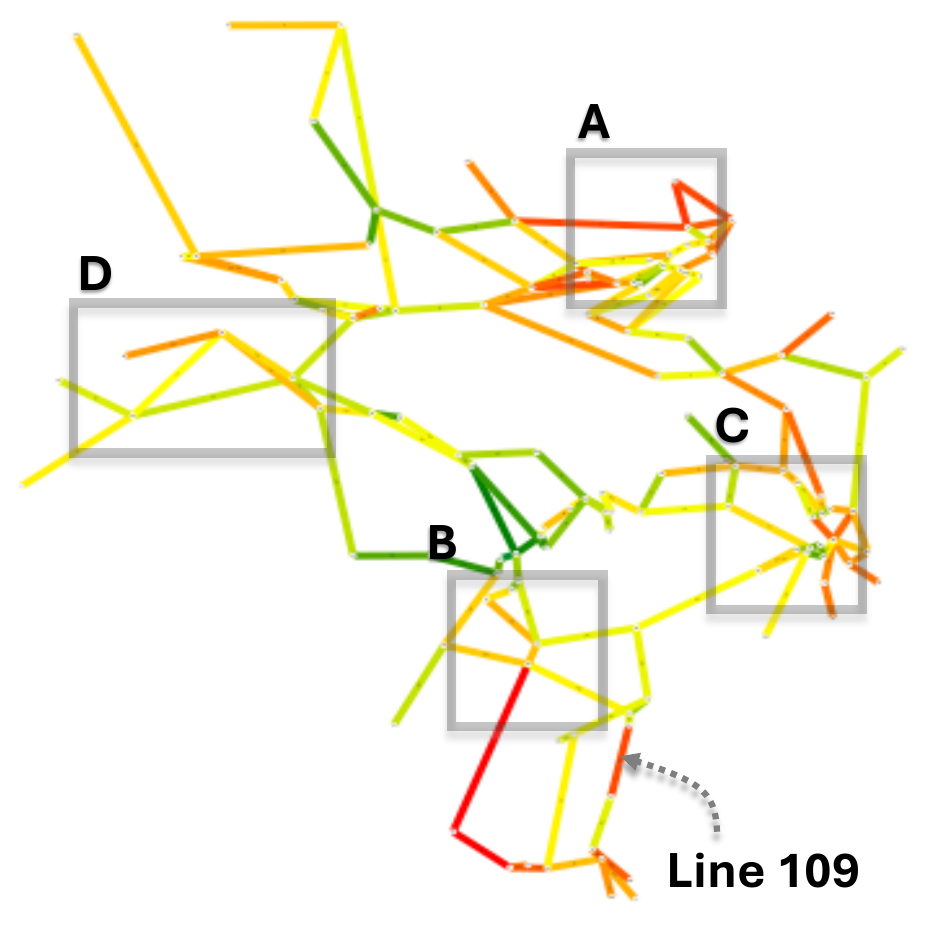}
         \vspace{-5mm}
         \caption{\small QRF. \vspace{-1mm}}
         \label{fig:prob_forecasting_QRF}
     \end{subfigure}
     \begin{subfigure}[b]{0.52\columnwidth}
     \centering
         \includegraphics[width=\columnwidth]{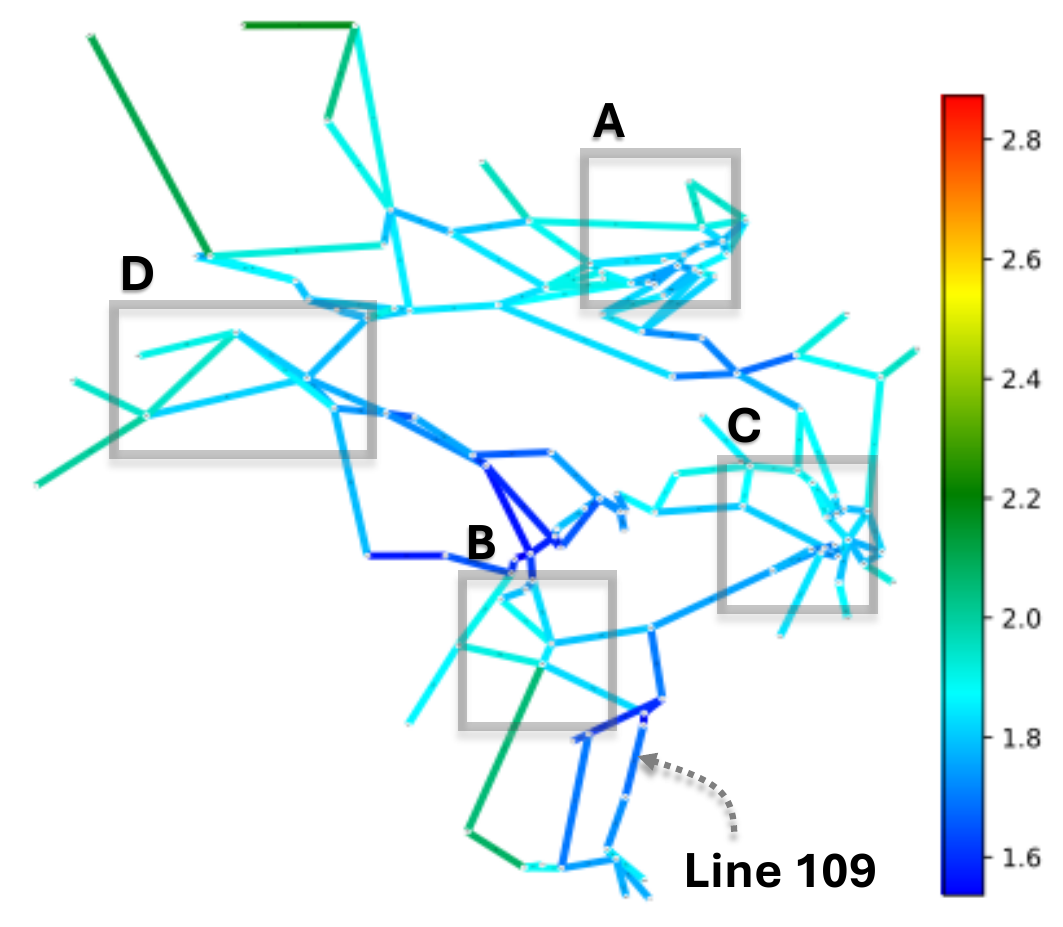}
         \vspace{-5mm}
         \caption{\small LGCLSTM.  \vspace{-1mm}}
         \label{fig:prob_forecasting_LGCLSTM}
     \end{subfigure}
     \caption{\small Heat maps of the average QS for each line in 2021 across TX-123BT. The gray dotted arrow points the line \#109 where the highest difference in QS between QRF to LGCLSTM is observed.}
     \label{fig:heat_map_QS_qrf_tste}\vspace{-3mm}
\end{figure}

In addition to the significant forecasting performance, LGCLSTM reduces the number of parameters by approximately 40\% and 93\% compared to GCLSTM and T-GCN, as shown in Table~\ref{table:param_numbers}. This is due to the broader message passing on the line graph, which captures extended spatial patterns with a single layer. In particular, although T-GCN has the highest number of parameters, it does not achieve the best results. This shows that increasing the complexity of the model does not necessarily improve the performance.

\subsubsection{Benefits of Network-Wide Consideration}
To demonstrate the benefits of incorporating spatial features in probabilistic DLR forecasting, we illustrate heat maps of QS for each line using test data across TX123BT in Fig.~\ref{fig:heat_map_QS_qrf_tste}. Note that the QS of each line is averaged over all evaluated prediction intervals in Table~\ref{table:simulation_results}. Specifically, we compare the performance of QRF and LGCLSTM to verify the importance of spatial information for accurate probabilistic forecasting.

In Fig.~\ref{fig:heat_map_QS_qrf_tste}, \textcolor{red}{red} indicates high QS (poorer performance), while \textcolor{blue}{blue} represents low QS (better performance). As can be seen, LGCLSTM generally exhibits a lower QS across the network compared to QRF. In particular, LGCLSTM achieves significant improvements in QS in regions A, B, C, and D, where neighboring buses are densely clustered. The improvements in these regions indicate the existence of a strong spatial correlation among transmission lines that can be effectively captured by LGCLSTM. Unlike QRF which treats each line independently and only captures temporal patterns, LGCLSTM leverages both temporal features and the network topology through line graph and extended-neighborhood message passing, that improves the forecasting performance.

{
\subsubsection{Prediction Interval of Probabilistic DLR Forecasting}

\begin{figure}[t]
     \centering
     \begin{subfigure}[b]{1\columnwidth}
     \centering
         \includegraphics[width=\columnwidth]{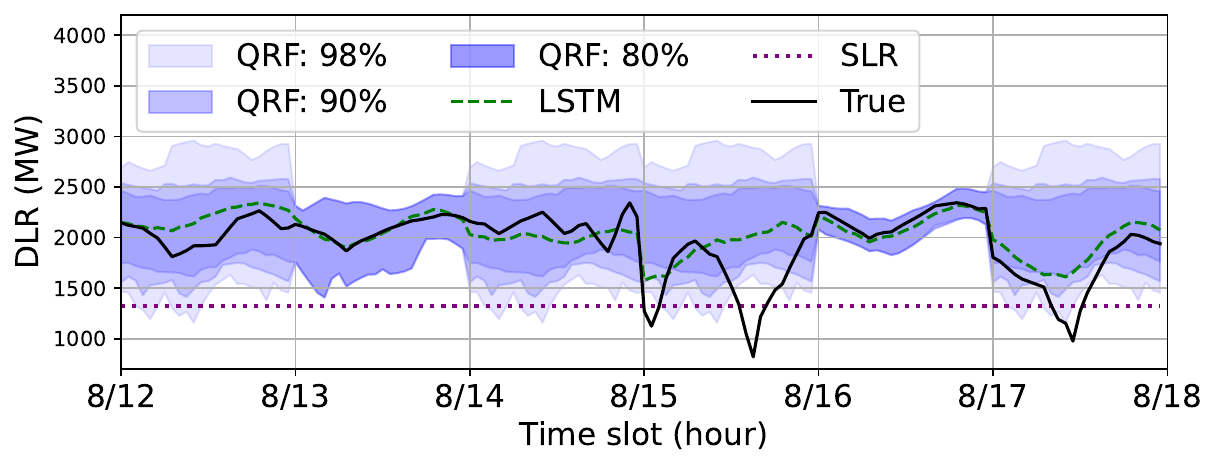}
         \vspace{-6.5mm}
         \caption{\small Probabilistic DLR forecasting using QRF.}
         \label{fig:prob_forecasting_QRF}
     \end{subfigure}
     \begin{subfigure}[b]{1\columnwidth}
     \centering
         \includegraphics[width=\columnwidth]{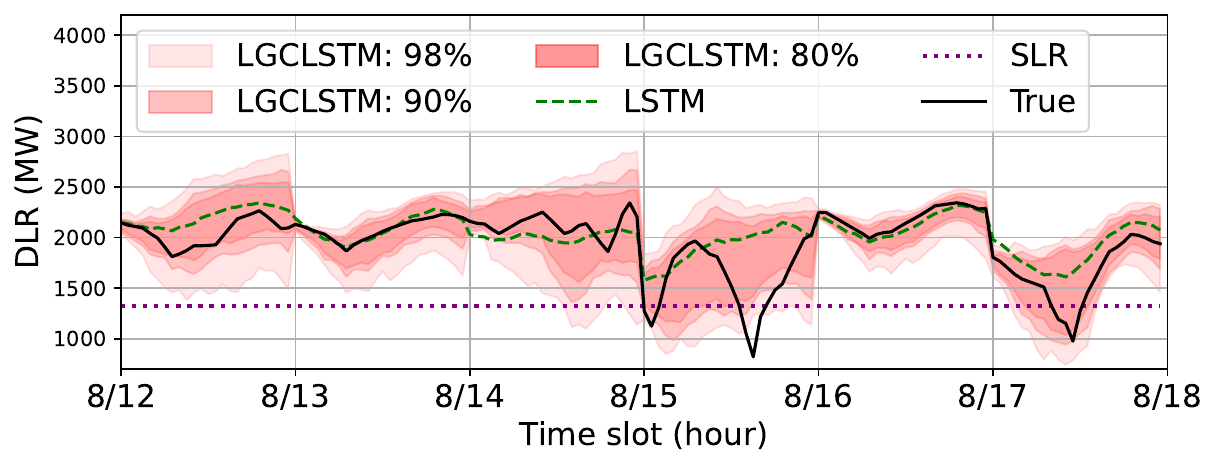}
         \vspace{-6.5mm}
         \caption{\small Probabilistic DLR forecasting using LGCLSTM.  \vspace{-1mm}}
         \label{fig:prob_forecasting_LGCLSTM}
     \end{subfigure}
     \caption{\small Probabilistic DLR forecasting for line \#109. \vspace{-4mm}}
     \label{fig:forecastings}
\end{figure}

The advantages of incorporating spatial features become even clearer when examining line-level behavior. To further investigate the forecasting characteristics, we compare deterministic and probabilistic DLR forecasts, SLR, and true DLR for a specific transmission line, as shown in Fig.~\ref{fig:forecastings}.
We select line~\#109, which exhibits the largest improvement in QS when transitioning from QRF to $\text{LGCLSTM}$. The analysis focuses on a one-week period in summer (August~12–18), when ambient temperatures are high and transmission lines are more susceptible to overheating

During the period from 8/15 to 8/16, the true DLR decreases sharply and even falls below the static line rating (SLR), indicating that SLR is not always conservative under adverse weather conditions. The deterministic forecast from LSTM fails to capture the details of true DLR and overestimates the available capacity, creating the potential risk of overloading. QRF generates wide prediction intervals, but still fails to enclose the true DLR at several points. In contrast, LGCLSTM closely follows both the magnitude and timing of the decline with substantially sharper intervals, demonstrating an improved ability to represent rapid variations in DLR.

Another drop occurs from 8/17 to 8/18, when all baselines (SLR, LSTM, and QRF) overestimate line capacity. LGCLSTM again captures the true DLR, reflecting the benefit of leveraging spatial features. Throughout both challenging periods, LGCLSTM produces bounds that remain close to the actual available capacity without unnecessary conservatism, which is consistent with the superior values of PINAW, IS and QS observed in Table~\ref{table:simulation_results}.
}

{\subsubsection{Error Analysis of DLR Forecasts and SLR} Fig.~\ref{fig:delta_capability} shows the percentage error  between the true DLR and four models (DLR forecasts and SLR). The DLR percentage error is defined as the percentage deviation from the true DLR, where positive values indicate forecasted DLR or SLR is higher than true DLR and negative values indicate forecasted DLR or SLR is lower than true DLR

\begin{figure}[t]
     \centering
         \includegraphics[width=1\columnwidth]{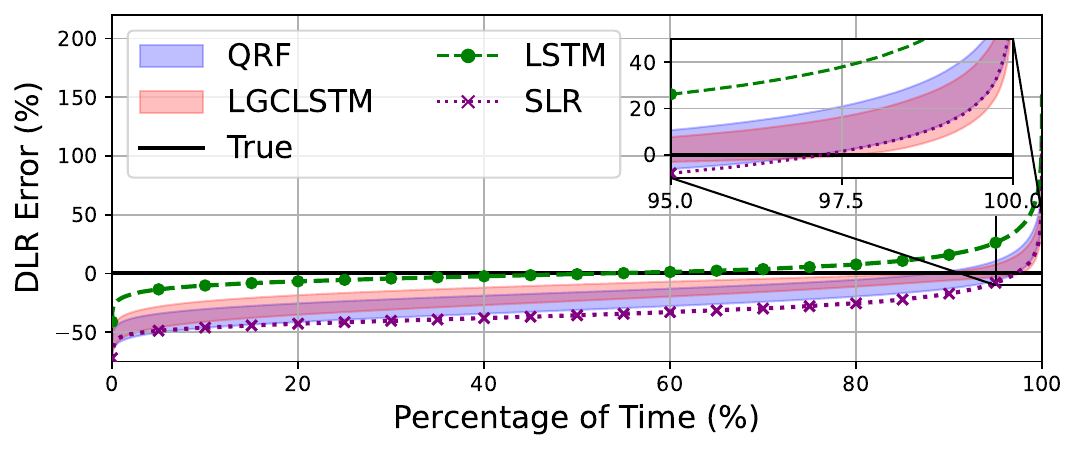}
         \vspace{-6mm}
     \caption{\small Comparisons of DLR percentage error (percentage deviation from true DLR; sorted in ascending order) of the entire network. \vspace{-4mm}}
     \label{fig:delta_capability}
\end{figure}
    
Although LSTM remains numerically close to zero, nearly half of its values are negative, indicating frequent overestimation that may lead to overload risk and unexpectedly high real-time redispatch costs. For the probabilistic models, the shaded bands represent the employed DLR derived from the 10\% and 1\% lower quantiles (corresponding to the 80\% and 98\% prediction intervals), since system operators use lower quantiles as conservative line limits to ensure thermal security. LGCLSTM stays closer to the true capacity than QRF and shows a narrower band, meaning that tightening the risk level leads to a smaller reduction in usable DLR. This is consistent with Table~\ref{table:simulation_results} and Fig.~\ref{fig:forecastings}, where LGCLSTM provides sharper yet reliable quantile forecasts.
    
In addition, although SLR is generally expected to pose little risk, Fig.~\ref{fig:delta_capability} shows that it still produces about 3\% overestimation events. While this 3\% rate is lower than the risk levels associated with the probabilistic models and the  overestimation frequency of LSTM, it does not necessarily translate into superior operational performance, as discussed next in Section~\ref{subsec:opertion_results}.

{
\vspace{-3mm}\subsection{Overall Performance in Grid Operation}\label{subsec:opertion_results}

\begin{table*}[t]
  \centering
  \captionsetup{justification=centering, labelsep=period, font=footnotesize, textfont=sc}
  \caption{\small Hourly average costs in 2021. The best and the second best results among the DLR forecasts are in \textbf{bold} and \underline{underlined}, respectively.\vspace{-2mm}}
  \label{table:cost_results}
  \begin{tabular}{c|l|c|ccccc|cc}
    \toprule
    \multicolumn{2}{c|}{\textbf{\makecell{Employed DLR\\for grid operation}}} 
    & \textbf{\makecell{DA cost\\(\$1k)}} 
    & \textbf{\makecell{Up RD\\(100MW)}}
    & \textbf{\makecell{Down RD\\(100MW)}}
    & \textbf{\makecell{Up RD\\cost (\$1k)}}
    & \textbf{\makecell{Down RD\\cost (\$1k)}}
    & \textbf{\makecell{RD cost\\(\$1k)}}
    & \textbf{\makecell{Total cost\\(\$1k)}}
    & \textbf{\makecell{CVaR cost\\(10\%, \$1k)}} \\
    \midrule\midrule
    \multicolumn{10}{c}{Hourly average total load: 38.11 GW (True), 39.38 GW (Pred)}\\
    \midrule\midrule
   Deterministic & LSTM & \textbf{1171.65}& 13.00 & 25.33 & 87.50 & \textbf{-27.33} & 60.17 & 1231.83 & 1838.86 \\
   \multirow{2}{*}{\makecell{Probabilistic\\$(Q_L = 0.10)$}}&QRF     & 1172.76  & \underline{12.96} & \underline{25.29} & \textbf{86.38} & -27.58 & \textbf{58.79} & \underline{1231.55} & \underline{1834.04} \\
    &$\text{LGCLSTM}$ & \underline{1172.44}  &$\textbf{12.95}$ & $\textbf{25.28}$ & \underline{86.46} & \underline{-27.50} & \underline{58.96} & \textbf{1231.40} & \textbf{1833.75} \\
    \midrule
    Deterministic & LSTM & \textbf{1171.65} & 13.00 & 25.33 & 87.50 & \textbf{-27.33} & 60.17 & 1231.83 & 1838.86 \\
    \multirow{2}{*}{\makecell{Probabilistic\\$(Q_L = 0.05)$}}&QRF     & 1173.72     & \underline{12.93} & \underline{25.27} & \textbf{85.68} & -27.81 & \textbf{57.87} & \underline{1231.59} & \underline{1832.17} \\
    &$\text{LGCLSTM}$ & \underline{1173.06} &$\textbf{12.92}$ & $\textbf{25.26}$ & \underline{85.92} & \underline{-27.64} & \underline{58.28} & \textbf{1231.33} & \textbf{1831.97} \\
    \midrule
    Deterministic & LSTM & \textbf{1171.65}  & 13.00 & 25.33 & 87.50 & \textbf{-27.33} & 60.17 & \underline{1231.83} & 1838.86 \\
    \multirow{2}{*}
    {\makecell{Probabilistic\\$(Q_L = 0.01)$}}&QRF     & 1176.39     & \underline{12.91} & \underline{25.25} & \textbf{84.13} & -28.44 & \textbf{55.69} & 1232.08 & \textbf{1829.44} \\
    &$\text{LGCLSTM}$ & \underline{1175.40}  &$\;\;\textbf{12.91}^*$ & \;\;$\textbf{25.25}^*$ & \underline{84.58} & \underline{-28.22} & \underline{56.36} & \textbf{1231.75} & \underline{1829.80} \\
    \midrule\midrule
    \multicolumn{2}{c|}{SLR }  & 1177.58 & 13.11 & 25.41    & 91.92    & -28.49   & 63.44  & 1241.02 & 1881.27 \\\midrule
    \multicolumn{2}{c|}{True DLR }  & 1171.99 & 12.91 & 25.24    & 86.34    & -27.33   & 59.01  & 1231.00 & 1832.27 \\
    \midrule\midrule
    \multicolumn{2}{c|}{Oracle (True DLR and load)}   & 1146.06 & 0.00    & 0.00    & 0.00     & 0.00     & 0.00     & 1146.06 & 1567.23 \\
    \bottomrule
  \end{tabular}
  \captionsetup{justification=raggedright, singlelinecheck=false, font=footnotesize, textfont=normalfont}
  \caption*{\;\;\quad\quad \small* indicates the value is lower before rounding. DA: Day-ahead, RD: Redispatch}
  \vspace{-5mm}
\end{table*}

We evaluate how different DLR forecasting models influence system operation. For all methods except Oracle (i.e., LSTM, QRF, LGCLSTM, SLR, and True DLR), the day-ahead (DA) problem is solved using load forecasts, while the real-time (RT) problem uses true load values. The True DLR case assumes perfect knowledge of line ratings in the DA stage, whereas Oracle assumes perfect knowledge of both DLR and load and represents the theoretical minimum operating cost. In the probabilistic approaches, the lower quantile of each prediction interval is used as the day-ahead line limit because it provides a conservative and operationally safe estimate of available capacity under DLR uncertainty. We highlight the lowest and second-lowest costs among the forecasting-based methods in \textbf{bold} and \underline{underline}, respectively, in Table~\ref{table:cost_results}. In addition, LGCLSTM achieves the best or second-best performance. All supporting figures in this section use $Q_L = 0.05$, since LGCLSTM with $Q_L = 0.05$ delivers the lowest total cost among all forecasting methods and quantile choices, and this setting most clearly reveals the operational trade-offs examined in this section.

\subsubsection{Day-Ahead Scheduling}

We begin by comparing the hourly average generation cost of DA scheduling (DA cost). The forecasting differences lead to different DA scheduling results. Since DA scheduling seeks to utilize inexpensive generators while satisfying forecasted DLRs, LSTM produces the lowest DA cost: its overestimated line ratings relax flow constraints and permit greater use of cheap generation (see Fig.~\ref{fig:DLR_mpe}). Notably, LSTM yields an even lower DA cost than the True DLR case, because its optimistic DLR predictions allows more aggressive use of inexpensive units. However, QRF provide conservative lower-quantile forecasts, which make tighter line limits and higher DA costs. By contrast, LGCLSTM achieves a balance between these. Although SLR exhibits less overestimation than the $Q_L = 0.01$ case, as shown in Fig.~\ref{fig:delta_capability}, it produces a higher DA cost because its static rating cannot exploit temporal variations of true DLR, leading to persistently conservative scheduling. These DA scheduling patterns directly influence the redispatch outcomes.

}

\begin{figure}[t]
     \centering
         \includegraphics[width=1\columnwidth]{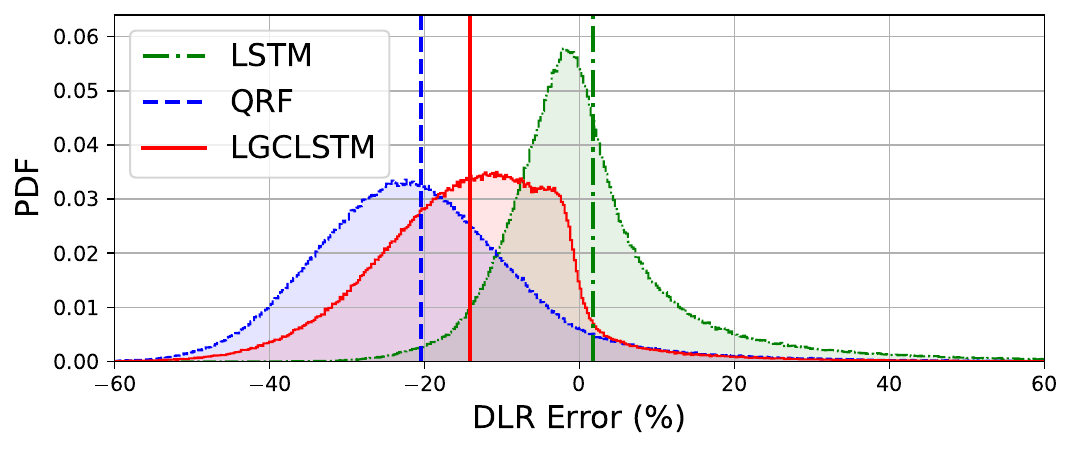}
         \label{fig:ERCOT_load}\vspace{-6mm}
     \caption{\small Distribution of hourly DLR percentage error of $Q_L = 0.05$ for DLR forecasts. Vertical lines represent the average (LSTM: 1.96\%, QRF: -20.42\%, LGCLSTM: -15.68\%). \vspace{-4mm}}
     \label{fig:DLR_mpe}
\end{figure}

{
\subsubsection{Real-Time Redispatch}

Since ERCOT's load forecast is positively biased, as shown in the second row of Table~\ref{table:cost_results}, downward redispatch dominates across all methods. However, the magnitude and cost of redispatch depend on how each model shapes the day-ahead schedule. LSTM requires the largest RT adjustments among forecasting-based approaches: its overestimated DLRs lead to aggressive day-ahead flows that become infeasible once true line limits are revealed, resulting in high upward redispatch cost. By contrast, conservative lower-quantile forecasts from probabilistic approaches avoid risky day-ahead schedules that would otherwise require costly real-time corrections. As a result, QRF attains the lowest total redispatch cost across all quantile levels due to its highly conservative lower quantile forecasts. LGCLSTM consistently achieves the second-lowest redispatch cost, and its redispatch cost remains much closer to QRF than to LSTM.

The True DLR case shows that even with perfect DLR information, load uncertainty alone can drive substantial redispatch. Interestingly, True DLR consistently yields higher redispatch cost than all probabilistic DLR approaches. This is because conservative lower quantile forecasts embed an operational safety margin that absorbs load forecast errors and suppresses extreme redispatch events more effectively than perfect DLR information alone.  Meanwhile, SLR performs worst in RT operation because its static rating fails to utilize hours with high actual DLR, leading to persistently conservative flows and frequent redispatch. Overall, LGCLSTM provides the most balanced RT performance across practical risk levels.}

\subsubsection{Total Cost and CVaR cost}

Now, we compare the total cost, which is the sum of DA cost and RD cost, and its CVaR cost of each method. LGCLSTM achieves the lowest cost among all forecasting-based methods for every quantile level (\$\textbf{1231.33}k). This reflects LGCLSTM’s ability to exploit spatio-temporal patterns of DLRs, in contrast to LSTM and QRF, which ignore spatial dependencies across the network.  Furthermore, the best overall performance is achieved at $Q_L = 0.05$. This demonstrates that moderate safety margins provide the most cost-effective operation, while overly conservative limits reduce efficiency. Thus, selecting an appropriate safety margin is crucial for economical operation. Probabilistic DLR forecasting enables this flexibility by allowing operators to tune the margin through the choice of prediction interval.

Among the forecasting-based methods, deterministic LSTM yields the highest CVaR due to frequent overestimation, while SLR exhibits the highest total and CVaR cost overall. Although Fig.~\ref{fig:delta_capability} shows that SLR exhibits a similar level of overestimation to the $Q_L = 0.01$, its total cost and CVaR remain significantly higher because SLR cannot capture the temporal variations of DLR and therefore cannot exploit periods of high true capacity. However, LGCLSTM maintains a low CVaR by achieving the lowest at $Q_L = 0.05$ (\$\textbf{1831.97}k) and $Q_L = 0.10$ (\$\textbf{1833.75}k). In the highly conservative $Q_L = 0.01$ setting, QRF achieves a slightly lower CVaR (\$\textbf{1829.44}k) than LGCLSTM due to its very tight limits from the lower quantile of wide prediction intervals. Interestingly, True DLR shows higher CVaR than LGCLSTM and QRF in this setting (\$1832.27k), since the additional safety margins from lower quantiles better absorb load forecasting errors by suppressing extreme redispatch.

\subsubsection{Curtailment Comparisons} 

\begin{table}[t]
  \centering
  \captionsetup{justification=centering, labelsep=period, font=footnotesize, textfont=sc}
  \caption{\small Hourly average curtailment in 2021. The worst result is shown in \textcolor{red}{\textbf{red}}.\vspace{-2mm}}
  \label{table:curtailment}
  \begin{tabular}{c|l|c|c}
    \toprule
    \multicolumn{2}{c|}{\textbf{\makecell{Employed DLR\\for grid operation}}} 
    & \textbf{\makecell{DA Cur. (MWh)}} 
    & \textbf{\makecell{RT Cur. (MWh)}}\\
    \midrule\midrule
    \multirow{2}{*}{\makecell{Probabilistic\\$(Q_L = 0.10)$}}
      & QRF               & 76.31    & 124.36 \\
      & $\text{LGCLSTM}$ & {75.97} & {124.37} \\
    \midrule
    \multirow{2}{*}{\makecell{Probabilistic\\$(Q_L = 0.05)$}}
      & QRF               & 77.18    & 124.38 \\
      & $\text{LGCLSTM}$ & {76.59} & {124.36} \\
    \midrule
    \multirow{2}{*}{\makecell{Probabilistic\\$(Q_L = 0.01)$}}
      & QRF               & 76.58    & 124.35 \\
      & $\text{LGCLSTM}$ & {76.78} & {124.35} \\
    \midrule
    \multicolumn{2}{c|}{LSTM}  
      & 76.41 & 124.40 \\
    \midrule

    \multicolumn{2}{c|}{SLR}  
      & \textcolor{red}{79.38} & \textcolor{red}{131.34} \\
    \midrule

    \multicolumn{2}{c|}{True DLR}  
      & 76.38 & 124.34 \\
    \midrule\midrule

    \multicolumn{2}{c|}{Oracle (True DLR and load)}   
      & \multicolumn{2}{c}{120.34} \\
    \bottomrule
  \end{tabular}
  \captionsetup{justification=raggedright, singlelinecheck=false, font=footnotesize, textfont=normalfont}
  \caption*{\small Cur.: Curtailment, DA: Day-ahead, RT: Real-time}
  \vspace{-6mm}
\end{table}

 As shown in Table~\ref{table:curtailment}, all forecasting-based methods exhibit nearly identical curtailment in both the day-ahead and real-time stages. In contrast, SLR shows the highest curtailment even when its overestimation frequency is similar to the highly conservative $Q_L = 0.01$ setting in Fig.~\ref{fig:delta_capability}. This is because SLR cannot track the temporal variations of DLR and therefore does not take advantage of periods with higher capacity. Thus, DLR forecasting is necessary for efficient utilization of renewables.

\subsubsection{Hourly Operational Performance in Months}\label{subsec:monthly_opertion_results}

Fig.~\ref{fig:Monthly_cost_and_error} compares the hourly total cost differences of QRF, True DLR, and LGCLSTM with SLR of each month throughout~2021. When the hourly load forecasting error in a month is low, the cost gap between SLR and the other methods is small. However, during months with high load uncertainty (e.g., February and September), the performance gap widens substantially, indicating that SLR is highly sensitive to load forecast errors. In particular, in September, the True DLR case becomes slightly more expensive than LGCLSTM, suggesting that conservative probabilistic limits can better absorb load uncertainty than perfect DLR information. Overall, these monthly results demonstrate that DLR forecasting mitigates the operational impact of load forecast errors more effectively than SLR.

\begin{figure}[t]
     \centering
         \includegraphics[width=1\columnwidth]{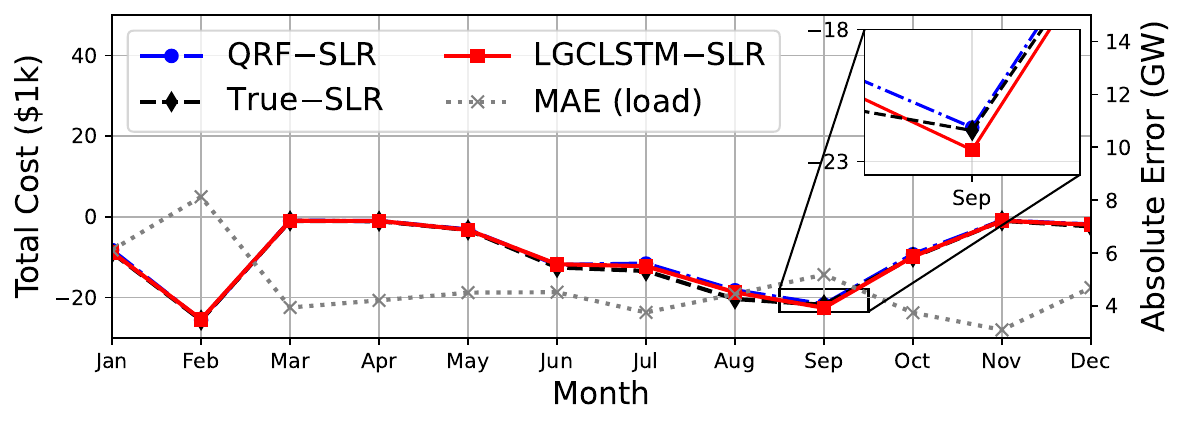}
         \label{fig:Monthly_cost_and_error}\vspace{-7mm}
     \caption{\small Total cost difference of $Q_L=0.05$ and load forecast error. MAE represents hourly mean absolute error of each month.\vspace{-5mm}} 
     \label{fig:Monthly_cost_and_error}
\end{figure}

\section{Conclusion}\label{sec:conclusion}

In this paper, we proposed a network-wide probabilistic DLR forecasting model called line graph convolutional LSTM (LGCLSTM), which integrates line graph convolutional networks into LSTM to exploit spatial and temporal information. By employing multi-hop message passing on the line graph, LGCLSTM captures extended spatial correlations. Extensive simulations on the Texas 123-bus backbone transmission system demonstrate that LGCLSTM outperforms all baselines in terms of reliability and sharpness while using the fewest parameters. We further evaluated the system operational impact of LGCLSTM by integrating its DLR forecasts into a day-ahead dispatch scheduling and a real-time redispatch framework. The results show that LGCLSTM achieves the least amount of redispatch and the lowest total cost among all forecasting-based baselines, while avoiding the excessive renewable curtailment observed under SLR due to its inability to track temporal variations of DLR. For future work, we plan to integrate LGCLSTM with advanced grid operations, such as security-constrained unit commitment or market operations.

\balance
\bibliographystyle{IEEEtran}
\bibliography{reference.bib}

@article{kipf2016semi,
  title={{Semi-supervised classification with graph convolutional networks}},
  author={Kipf, Thomas N and Welling, Max},
  journal={Int. Conf. Learn. Represent.},
  year={2017}
}

@article{lu2025synthetic,
  title={{A Synthetic Texas Power System with Time-Series Weather-Dependent Spatiotemporal Profiles}},
  author={Lu, Jin and others},
  journal={Sustainable Energy, Grids and Networks},
  pages={101774},
  year={2025},
  publisher={Elsevier}
}

@inproceedings{
	loshchilov2018decoupled,
	title={{Decoupled Weight Decay Regularization}},
	author={Loshchilov, Ilya and Hutter, Frank},
	booktitle={International Conference on Learning Representations},
	year={2019},
}

@article{gao2023day,
  title={{Day-ahead dynamic thermal line rating forecasting and power transmission capacity calculation based on ForecastNet}},
  author={Gao, Zhengnan and others},
  journal={Electric Power Systems Research},
  volume={220},
  pages={109350},
  year={2023},
  publisher={Elsevier}
}

@article{zhao2019t,
  title={{T-GCN: A temporal graph convolutional network for traffic prediction}},
  author={Zhao, Ling and others},
  journal={IEEE Trans. Intell. Transp. Syst.},
  volume={21},
  number={9},
  year={2019},
  publisher={IEEE}
}

@article{simeunovic2021spatio,
  title={{Spatio-temporal graph neural networks for multi-site PV power forecasting}},
  author={Simeunovi{\'c}, Jelena and others},
  journal= {IEEE Trans. Sustainable Energy},
  year={2021},
  publisher={IEEE}
}

@standard{IEEE738_2012,
  title        = {{IEEE Standard for Calculating the Current-Temperature Relationship of Bare Overhead Conductors}},
  institution  = {IEEE},
  number       = {738-2012},
  year         = {2012},
}

@article{douglass2019review,
  title={{A review of dynamic thermal line rating methods with forecasting}},
  author={Douglass, Dale A and others},
  journal={IEEE Transactions on Power Delivery},
  year={2019},
  publisher={IEEE}
}

@article{fernandez2016review,
  title={{Review of dynamic line rating systems for wind power integration}},
  author={Fernandez, E and others},
  journal={Renewable and Sustainable Energy Reviews},
  year={2016},
  publisher={Elsevier}
}

@article{ahmadi2023decomposition,
  title={{Decomposition-Based Stacked Bagging Boosting Ensemble for Dynamic Line Rating Forecasting}},
  author={Ahmadi, Amirhossein and others},
  journal={IEEE Transactions on Power Delivery},
  volume={38},
  number={5},
  pages={2987--2997},
  year={2023},
  publisher={IEEE}
}

@inproceedings{molinar2019ampacity,
  title={{Ampacity forecasting: an approach using Quantile Regression Forests}},
  author={Molinar, Gabriela and others},
  booktitle={2019 IEEE Power \& Energy Society Innovative Smart Grid Technologies Conference (ISGT)},
  year={2019},
  organization={IEEE}
}

@article{dupin2019optimal,
  title={{Optimal dynamic line rating forecasts selection based on ampacity probabilistic forecasting and network operators’ risk aversion}},
  author={Dupin, Romain and others},
  journal={IEEE Transactions on Power Systems},
  volume={34},
  number={4},
  year={2019},
  publisher={IEEE}
}

@article{viafora2020chance,
  title={{Chance-constrained optimal power flow with non-parametric probability distributions of dynamic line ratings}},
  author={Viafora, Nicola  and others},
  journal={International Journal of Electrical Power \& Energy Systems},
  volume={114},
  year={2020},
  publisher={Elsevier}
}

@article{madadi2019dynamic,
  title={{Dynamic line rating forecasting based on integrated factorized Ornstein--Uhlenbeck processes}},
  author={Madadi, Sajad and Mohammadi-Ivatloo, Behnam and Tohidi, Sajjad},
  journal={IEEE Transactions on Power Delivery},
  volume={35},
  number={2},
  year={2019},
  publisher={IEEE}
}

@article{song2024graph,
  title={{Graph-Based Large Scale Probabilistic PV Power Forecasting Insensitive to Space-Time Missing Data}},
  author={Song, Keunju and Kim, Minsoo and Kim, Hongseok},
  journal={IEEE Transactions on Sustainable Energy},
  year={2024},
  publisher={IEEE}
}

@article{sun2022spatio,
  title={{Spatio-temporal weather model-based probabilistic forecasting of dynamic thermal rating for overhead transmission lines}},
  author={Sun, Xiaorong and Jin, Chenhao},
  journal={International Journal of Electrical Power \& Energy Systems},
  year={2022},
  publisher={Elsevier}
}

@article{li2022integrated,
  title={{An integrated missing-data tolerant model for probabilistic PV power generation forecasting}},
  author={Li, Qiaoqiao and others},
  journal={IEEE Trans. on Power Syst.},
  year={2022},
  publisher={IEEE}
}

@article{wang2019probabilistic,
  title={{Probabilistic individual load forecasting using pinball loss guided LSTM}},
  author={Wang, Yi and others},
  journal={Applied Energy},
  volume={235},
  pages={10--20},
  year={2019},
  publisher={Elsevier}
}

@inproceedings{chen2020measuring,
  title={{Measuring and relieving the over-smoothing problem for graph neural networks from the topological view}},
  author={Chen, Deli and others},
  booktitle={Proceedings of the AAAI conference on artificial intelligence},
  volume={34},
  number={04},
  year={2020}
}

@article{liu2022topology,
  title={{Topology-aware graph neural networks for learning feasible and adaptive AC-OPF solutions}},
  author={Liu, Shaohui and Wu, Chengyang and Zhu, Hao},
  journal={IEEE Transactions on Power Systems},
  volume={38},
  number={6},
  pages={5660--5670},
  year={2022},
  publisher={IEEE}
}

@article{liao2021review,
  title={{A review of graph neural networks and their applications in power systems}},
  author={Liao, Wenlong and others},
  journal={J. Mod. Power Syst. Clean Energy},
  year={2021},
  publisher={SGEPRI}
}

@inproceedings{gilmer2017neural,
  title={{Neural message passing for quantum chemistry}},
  author={Gilmer, Justin and others},
  booktitle={International conference on machine learning},
  year={2017},
  organization={PMLR}
}

@article{velivckovic2017graph,
  title={Graph attention networks},
  author={Veli{\v{c}}kovi{\'c}, Petar and Cucurull, Guillem and Casanova, Arantxa and Romero, Adriana and Lio, Pietro and Bengio, Yoshua},
  journal={arXiv preprint arXiv:1710.10903},
  year={2017}
}

@inproceedings{li2021training,
  title={Training graph neural networks with 1000 layers},
  author={Li, Guohao and others},
  booktitle={International conference on machine learning},
  year={2021},
  organization={PMLR}
}

@article{meinshausen2006quantile,
  title={Quantile regression forests.},
  author={Meinshausen, Nicolai and Ridgeway, Greg},
  journal={Journal of machine learning research},
  volume={7},
  number={6},
  year={2006}
}

@article{kirilenko2020risk,
  title={{Risk-averse stochastic dynamic line rating models}},
  author={Kirilenko, Aleksei and others},
  journal={IEEE Transactions on Power Systems},
  volume={36},
  number={4},
  year={2020},
  publisher={IEEE}
}

@article{aznarte2016dynamic,
  title={{Dynamic line rating using numerical weather predictions and machine learning: A case study}},
  author={Aznarte, Jose L and Siebert, Nils},
  journal={IEEE Transactions on Power Delivery},
  volume={32},
  number={1},
  pages={335--343},
  year={2016},
  publisher={IEEE}
}

@article{saatloo2021hierarchical,
  title={Hierarchical extreme learning machine enabled dynamic line rating forecasting},
  author={Saatloo, Amin Mansour and others},
  journal={IEEE Systems Journal},
  year={2021},
  publisher={IEEE}
}

@inproceedings{martinez2024dynamic,
  title={{Dynamic line rating forecasting using recurrent neural networks}},
  author={Martinez, Roberto Fernandez and others},
  booktitle={2024 18th International Conference on Probabilistic Methods Applied to Power Systems (PMAPS)},
  year={2024},
  organization={IEEE}
}

@article{van2021machine,
  title={{Machine learning for optimal power flows}},
  author={Van Hentenryck, Pascal},
  journal={Tutorials in Operations Research: Emerging Optimization Methods and Modeling Techniques with Applications},
  pages={62--82},
  year={2021},
  publisher={Informs}
}

@article{bhattarai2018improvement,
  title={{Improvement of transmission line ampacity utilization by weather-based dynamic line rating}},
  author={Bhattarai, Bishnu P and others},
  journal={IEEE Transactions on Power Delivery},
  volume={33},
  number={4},
  pages={1853--1863},
  year={2018},
  publisher={IEEE}
}

@article{dvorkin2025regression,
  title={{Regression equilibrium in electricity markets}},
  author={Dvorkin, Vladimir},
  journal={IEEE Transactions on Energy Markets, Policy and Regulation},
  year={2025},
  publisher={IEEE}
}

@article{pritchard2010single,
  title={{A single-settlement, energy-only electric power market for unpredictable and intermittent participants}},
  author={Pritchard, Geoffrey and others},
  journal={Oper. Res.},
  year={2010},
  publisher={INFORMS}
}

@article{zhou2017graph,
  title={{Graph convolution: A high-order and adaptive approach}},
  author={Zhou, Zhenpeng and Li, Xiaocheng},
  journal={arXiv preprint arXiv:1706.09916},
  year={2017}
}

@online{ferc2024dlr,
  title        = {Implementation of Dynamic Line Ratings},
  author       = {{FERC}},
  year         = {2024},
  organization = {Federal Register}
}

@online{ferc2024transmission,
  title        = {Fact Sheet: Building for the Future Through Electric Regional Transmission Planning and Cost Allocation},
  author       = {{FERC}},
  year         = {2024},
  organization = {FERC}
}

@manual{ERCOTNodalProtocols,
  title        = {ERCOT Nodal Protocols},
  organization = {{ERCOT}},
  month        = {March 28},
  year         = {2025},
}

@article{kim2024probabilistic,
  title={{Probabilistic Dynamic Line Rating Forecasting with Line Graph Convolutional LSTM}},
  author={Kim, Minsoo and others},
  journal={arXiv preprint arXiv:2411.12963},
  year={2024}
}

@techreport{phillips2021forecasting,
  title={{Forecasting dynamic line rating with spatial variation considerations}},
  author={Phillips, Tyler Bennett and others},
  year={2021},
  institution={Idaho National Lab.(INL)}
}

@book{harary2018graph,
  title={{Graph theory (on Demand Printing of 02787)}},
  author={Harary, Frank},
  year={2018},
  publisher={CRC Press}
}
\end{document}